\documentclass[aip,graphicx]{revtex4-2}
\usepackage{graphicx}
\usepackage{amsmath}
\usepackage{amsfonts}
\usepackage{amssymb}

\usepackage{comment}

\usepackage{color} 
\usepackage[normalem]{ulem} 
\usepackage{tikz-cd} 

\usepackage{subcaption} 

\newlength{\mm}
\newlength{\cm}
\newcommand{\bef}[1]{\textcolor{red}{\sout{#1}}}
\renewcommand{\bef}[1]{} 
\newcommand{\befEq}[1]{\textcolor{red}{#1}}
\renewcommand{\befEq}[1]{} 
\newcommand{\aft}[1]{\textcolor{blue}{#1}}
\renewcommand{\aft}[1]{#1} 

\providecommand\bcdot{\boldsymbol{\cdot}}

\newcommand{\ud}{\mathrm{d}}

\newcommand\mnewcommand[1]{%
\let#1\relax \newcommand#1 }

\newcommand{\F}[1]{\overline{#1}}

\newcommand{\ie}{\textit{i.e.,}}
\newcommand{\eg}{\textit{e.g.,}}

\newcommand{\etal}{\textit{et al.}}

\newcommand{\mbigskip}{\vspace{2mm}}

\mnewcommand{\mysubsubsection}{\subsubsection}
\mnewcommand{\comentar}[1]{}

\mnewcommand{\Nx}{N_\x}
\mnewcommand{\Ny}{N_\y}
\mnewcommand{\Nz}{N_\z}
\mnewcommand{\Nm}{N_m}
\mnewcommand{\Ns}{N_s}
\mnewcommand{\complconj}[1]{#1^{*}}
\mnewcommand{\mvbrack}[1]{\left[ #1 \right]}
\mnewcommand{\step}{\Delta}
\mnewcommand{\dt}{\step t}
\mnewcommand{\traspose}{^{T}}
\mnewcommand{\avgtime}[1]{\left< #1 \right>}
\mnewcommand{\avg}[1]{\overline{#1}}
\mnewcommand{\mcdot}{\bcdot}
\mnewcommand{\mnabla}{\nabla}

\mnewcommand{\real}{\mathbb{R}}
\mnewcommand{\complex}{\mathbb{C}}
\mnewcommand{\imag}{\mathbb{I}}

\mnewcommand{\foutrans}[1]{\hat{#1}}
\mnewcommand{\modetrans}[1]{\tilde{#1}}

\mnewcommand{\timeintparam}{\kappa}

\mnewcommand{\x}{x}
\mnewcommand{\y}{y}
\mnewcommand{\z}{z}

\mnewcommand{\vels}{u_s}
\mnewcommand{\uvel}{u}
\mnewcommand{\vvel}{v}
\mnewcommand{\wvel}{w}

\mnewcommand{\vortx}{\omega_\x}
\mnewcommand{\vorty}{\omega_\y}
\mnewcommand{\vortz}{\omega_\z}

\mnewcommand{\facevel}{[\velh]_\nface}

\mnewcommand{\flux}{f}
\mnewcommand{\Dx}{\Delta x}
\mnewcommand{\Dy}{\Delta \y}
\mnewcommand{\Dz}{\Delta z}

\mnewcommand{\isvector}[1]{\boldsymbol{#1}}
\mnewcommand{\istensor}[1]{\mathsf{#1}}

\mnewcommand{\va}{\isvector{a}}
\mnewcommand{\vb}{\isvector{b}}
\mnewcommand{\vc}{\isvector{c}}
\mnewcommand{\vd}{\isvector{d}}
\mnewcommand{\sca}{\phi}
\mnewcommand{\scafield}{\isvector{\phi}}
\mnewcommand{\scafieldc}{\scafield_{c}}

\mnewcommand{\vel}{\isvector{u}}
\mnewcommand{\velv}{\isvector{v}}
\mnewcommand{\velw}{\isvector{w}}
\mnewcommand{\normal}{\isvector{n}}

\mnewcommand{\vela}{\vel_{a}}
\mnewcommand{\mua}{\mu_{a}}

\mnewcommand{\velhcsol}{\velhc^{sol}} 

\mnewcommand{\Rh}{R_{h}}

\mnewcommand{\veluc}{\isvector{u}_1}
\mnewcommand{\velvc}{\isvector{u}_2}
\mnewcommand{\velwc}{\isvector{u}_3}

\mnewcommand{\basis}{\isvector{w}}

\mnewcommand{\tensor}{\istensor{T}}
\mnewcommand{\Identity}{\istensor{I}}

\mnewcommand{\bodyforce}{\isvector{f}}
\mnewcommand{\velh}{\vel_s}
\mnewcommand{\velhv}{\velv_s}
\mnewcommand{\velhw}{\velw_s}
\mnewcommand{\tildevelh}{\tilde{\vel}_s}
\mnewcommand{\velhc}{\vel_c}
\mnewcommand{\velhcCB}[1]{\ifthenelse{\equal{#1}{}}{\velhc^{\ominus}}{\velhc^{#1,\ominus}}}
\mnewcommand{\velhcCBFree}[1]{\ifthenelse{\equal{#1}{}}{\velhc^{\oplus}}{\velhc^{#1,\oplus}}}
\mnewcommand{\velhcHF}[1]{\ifthenelse{\equal{#1}{}}{\velhc^{>}}{\velhc^{#1,>}}}
\mnewcommand{\velhcLF}[1]{\ifthenelse{\equal{#1}{}}{\velhc^{<}}{\velhc^{#1,<}}}
\mnewcommand{\velhsHF}[1]{\ifthenelse{\equal{#1}{}}{\velh^{>}}{\velh^{#1,>}}}
\mnewcommand{\velhsLF}[1]{\ifthenelse{\equal{#1}{}}{\velh^{<}}{\velh^{#1,<}}}
\mnewcommand{\vortd}{\vort_v}
\mnewcommand{\presh}{\isvector{p}_c}
\mnewcommand{\bodyforceh}{\bodyforce_c}
\mnewcommand{\diver}{\mnabla \cdot}
\mnewcommand{\lapl}{\mnabla^2}
\mnewcommand{\grad}{\mnabla}
\mnewcommand{\Ru}[1]{\isvector{R} \left( #1 \right)}
\mnewcommand{\Rud}[1]{\mathsfbi{R} \left( #1 \right)}
\mnewcommand{\vecnull}{\isvector{0}}
\mnewcommand{\vecnulls}{\vecnull_{s}}
\mnewcommand{\vecnullg}{\vecnull_{h}}
\mnewcommand{\vecnullc}{\vecnull_{c}}
\mnewcommand{\vecone}{\isvector{1}}
\mnewcommand{\veconec}{\vecone_{c}}
\mnewcommand{\vecones}{\vecone_{s}}
\mnewcommand{\veconeg}{\vecone_{h}}
\mnewcommand{\veconetresc}{\vecone_{3c}}

\mnewcommand{\velhg}{\vel_h}
\mnewcommand{\preshg}{\isvector{p}_h}

\mnewcommand{\dim}{3}  

\mnewcommand{\ud}{d}
\mnewcommand{\vort}{\isvector{\omega}}
\mnewcommand{\rot}[1]{\mnabla \times #1}
\mnewcommand{\selfinnerprod}[1]{\innerprod{#1}{#1}}
\mnewcommand{\innerprod}[2]{( #1 , #2 )}
\mnewcommand{\convective}[2]{C \left( #1 , #2 \right)}
\mnewcommand{\intvol}[1]{\int_{\Omega} #1 \ud \Omega}
\mnewcommand{\intsurf}[1]{\int_{\partial \Omega} #1 \ud S}

\mnewcommand{\nvc}{k}
\mnewcommand{\nedge}{v}
\mnewcommand{\axis}{i}
\mnewcommand{\nface}{f}
\mnewcommand{\cp}{c}
\mnewcommand{\cpA}{{c1}}
\mnewcommand{\cpB}{{c2}}
\mnewcommand{\Fedge}[1]{F_e ( #1 )}
\mnewcommand{\Fcell}[1]{F_f ( #1 )}
\mnewcommand{\Fvolume}[1]{F_c ( #1 )}

\mnewcommand{\mathsfbi}[1]{\mathsf{#1}}

\mnewcommand{\conv}{\mathsfbi{C}\left( \velh \right)}
\mnewcommand{\convg}{\mathsfbi{C}\left( \velhg \right)}
\mnewcommand{\convc}{\mathsfbi{C}_{c} \left( \velh \right)}
\mnewcommand{\convvc}{\mathsfbi{C}_{c}^{\dim d} \left( \velh \right)}
\mnewcommand{\convu}{\mathsfbi{C}_{u} \left( \velh \right)}
\mnewcommand{\convtraspose}{\mathsfbi{C}\traspose\left( \velh \right)}
\mnewcommand{\convgtraspose}{\mathsfbi{C}\traspose\left( \velhg \right)}
\mnewcommand{\convarg}[1]{\mathsfbi{C}\left( #1 \right)}
\mnewcommand{\convargtraspose}[1]{\mathsfbi{C}\traspose\left( #1 \right)}
\mnewcommand{\convutraspose}{\mathsfbi{C}_{u} \traspose\left( \velh \right)}
\mnewcommand{\convmat}{\mathsfbi{C}}

\DeclareRobustCommand{\spreg}[2]{\ifthenelse{\equal{#2}{}}{\convmat_{#1}}{\convmat_{#1}\left( #2 \right)}}
\mnewcommand{\velauxc}{\velv_c}
\mnewcommand{\velauxWc}{\velw_c}

\mnewcommand{\velhcmode}[1]{\ifthenelse{\equal{#1}{+}}{\velhcHF{}}{\ifthenelse{\equal{#1}{-}}{\velhcLF{}}{\ifthenelse{\equal{#1}{=}}{\velhc}{}}}}
\mnewcommand{\velhsmode}[1]{\ifthenelse{\equal{#1}{+}}{\velhsHF{}}{\ifthenelse{\equal{#1}{-}}{\velhsLF{}}{\ifthenelse{\equal{#1}{=}}{\velh}{}}}}

\mnewcommand{\velhctype}[2]{\ifthenelse{\equal{#2}{f}}{\overline{\velhcmode{#1}}}{\ifthenelse{\equal{#2}{r}}{\left(\velhcmode{#1}\right)^\prime}{\ifthenelse{\equal{#2}{v}}{\velhcmode{#1}}{}}}}
\mnewcommand{\velhstype}[2]{\ifthenelse{\equal{#2}{f}}{\overline{\velhsmode{#1}}}{\ifthenelse{\equal{#2}{r}}{\left(\velhsmode{#1}\right)^\prime}{\ifthenelse{\equal{#2}{v}}{\velhsmode{#1}}{}}}}

\mnewcommand{\triadic}[6]{\left(\velhctype{#1}{#2}\right)\traspose \convarg{\velhstype{#3}{#4}} \velhctype{#5}{#6}}

\mnewcommand{\OP}{J}
\mnewcommand{\OPh}{\mathsfbi{\OP}}
\mnewcommand{\OPc}{{\cal \OP}}
\mnewcommand{\diff}{\mathsfbi{D}}
\mnewcommand{\diffg}{\mathsfbi{D}}
\mnewcommand{\diffc}{\diff_{\cp}}
\mnewcommand{\diffvc}{\diffc^{\dim d}}
\mnewcommand{\diffu}{\diff_{u}}
\mnewcommand{\dive}{\mathsfbi{M}}
\mnewcommand{\dives}{\dive_s}
\mnewcommand{\diveg}{\mathsfbi{M}_h}
\mnewcommand{\divescaf}{\dive_{sf}}
\mnewcommand{\divevecf}{\dive_{vf}}
\mnewcommand{\graddc}{\mathsfbi{G}_c}
\mnewcommand{\graddg}{\mathsfbi{G}_h}
\mnewcommand{\gradd}{\XaviAlex{\graddg}}
\mnewcommand{\lapld}{\mathsfbi{L}}
\mnewcommand{\lapldc}{\lapld_{\cp}}
\mnewcommand{\vcvects}{\mathsfbi{\Omega}_s}
\mnewcommand{\pseudovcvects}{\tilde{\mathsfbi{\Omega}}_s}
\mnewcommand{\vcvectg}{\mathsfbi{\Omega}_h}
\mnewcommand{\vcvectc}{\mathsfbi{\Omega}_{\cp}}
\mnewcommand{\vcvectv}{\mathsfbi{\Omega}_v}
\mnewcommand{\tildevcvectv}{\widetilde{\mathsfbi{\Omega}}_v}
\mnewcommand{\tildevcvects}{\widetilde{\mathsfbi{\Omega}}_s}
\mnewcommand{\vcvectvc}{\vcvectc^{\dim d}} 
\mnewcommand{\vcvect}{\mathsfbi{\Omega}}
\mnewcommand{\nullmat}{\mathsfbi{0}}
\mnewcommand{\normd}[1]{|| #1 ||}
\mnewcommand{\rotd}{\mathsfbi{R}}
\mnewcommand{\graddscaf}{\gradd_{sf}}
\mnewcommand{\graddvecf}{\gradd_{vf}}
\mnewcommand{\graddx}{\gradd_{\x}}
\mnewcommand{\graddy}{\gradd_{\y}}
\mnewcommand{\graddz}{\gradd_{\z}}
\mnewcommand{\graddd}[1]{\ifthenelse{\equal{#1}{1}}{\graddx}{\ifthenelse{\equal{#1}{2}}{\graddy}{\ifthenelse{\equal{#1}{3}}{\graddz}{\gradd_{x_i}}}}}
\mnewcommand{\graddprod}[2]{\graddd{#1}\traspose \vcvect \graddd{#2}}

\mnewcommand{\fluxh}[2]{T_{#1}\left({#2}\right)}

\mnewcommand{\twoD}{two-dimensional}
\mnewcommand{\threeD}{three-dimensional}
\mnewcommand{\TwoD}{Two-dimensional}
\mnewcommand{\ThreeD}{Three-dimensional}
\mnewcommand{\biD}{\twoD~}
\mnewcommand{\triD}{\threeD~}
\mnewcommand{\BiD}{\TwoD~}
\mnewcommand{\TriD}{\ThreeD~}
\mnewcommand{\biandtriD}{two- and \threeD~}
\mnewcommand{\BiandtriD}{Two- and \threeD~}
\mnewcommand{\Dim}[1]{\ifthenelse{\equal{#1}{2}}{\twoD}{\ifthenelse{\equal{#1}{3}}{\threeD}{KK}}}

\mnewcommand{\inttypeoftext}{\mathsf}
\mnewcommand{\kinener}{\inttypeoftext{E}}
\mnewcommand{\enstrophy}{\inttypeoftext{\mathcal{E}}}
\mnewcommand{\helicity}{\inttypeoftext{H}}
\mnewcommand{\vorthelicity}{\helicity_{\vort}}
\mnewcommand{\palinstrophy}{\inttypeoftext{P}}

\mnewcommand{\helicityd}{\helicity_c}
\mnewcommand{\enstrophyd}{\enstrophy_c}

\mnewcommand{\vvlength}{m}
\mnewcommand{\pvlength}{n}
\mnewcommand{\evlength}{e}

\mnewcommand{\Isc}{\Gamma}
\mnewcommand{\Ics}{\Isc\traspose}

\mnewcommand{\Scs}{\Gamma_{c \rightarrow s}}
\mnewcommand{\Ssc}{\Gamma_{s \rightarrow c}}

\mnewcommand{\Sscal}{\Pi_{c \rightarrow s}}
\mnewcommand{\Sthreescal}{\Pi}
\mnewcommand{\NormalVect}[1]{\ifthenelse{\equal{#1}{}}{\istensor{N}_{s}}{\istensor{N}_{s,#1}}}

\mnewcommand{\Correction}{\istensor{P}}
\mnewcommand{\PseudoCorrection}{\tilde{\Correction}}
\mnewcommand{\kernel}[1]{Ker \left( #1 \right)}

\mnewcommand{\Ivs}{\Psi}
\mnewcommand{\Isv}{\Psi\traspose}

\mnewcommand{\OIsc}{\Upsilon}
\mnewcommand{\OIcs}{\Pi}

\mnewcommand{\order}{o}
\mnewcommand{\coarsemesh}{i}
\mnewcommand{\sizevc}{{V}}
\mnewcommand{\error}{\epsilon}

\mnewcommand{\convH}{\text{(Conv)}_{\helicityd}}
\mnewcommand{\diffH}{\text{(Diff)}_{\helicityd}}
\mnewcommand{\presH}{\text{(Pres)}_{\helicityd}}

\mnewcommand{\Filter}[1]{\overline{#1}}
\mnewcommand{\FilterFVM}[1]{\overline{#1}^{FV}}
\mnewcommand{\FilterLength}{\delta}
\mnewcommand{\ExplFilter}[1]{\widetilde{#1}}
\mnewcommand{\ExplFilterLength}{\tilde{\delta}}
\mnewcommand{\DiscrFilter}[1]{\istensor{F}{#1}}
\mnewcommand{\DiscrResidual}[1]{\istensor{R}{#1}}
\mnewcommand{\HONS}{\mathsf{HO}}

\mnewcommand{\q}{\isvector{q}}

\mnewcommand{\FVMFilter}[1]{\widetilde{#1}}

\newcommand{\As}{\istensor{A}_{s}}

\newcommand{\diffm}{\istensor{\Lambda}_s}
\newcommand{\pdiffm}{\tilde{\istensor{\Lambda}}_s}
\newcommand{\diffmc}{\istensor{\Lambda}_c} 
\newcommand{\pdiffmc}{\tilde{\istensor{\Lambda}}_c} 
\newcommand{\Tcs}{\istensor{T}_{cs}}
\newcommand{\Tsc}{\istensor{T}_{sc}}
\newcommand{\DX}{\istensor{\Delta}_s}

\newcommand{\halving}[1]{\frac{#1}{2}}
\newcommand{\half}{\halving{1}}

\newcommand{\Flength}{\Xavi{\delta}}

\renewcommand{\P}{P}
\newcommand{\Q}{Q}
\newcommand{\R}{R}

\newcommand{\subnut}{t}
\newcommand{\nut}{\nu_{\subnut}}

\newcommand{\pseudovar}[1]{\hat{#1}}

\newcommand{\G}{\istensor{G}}
\renewcommand{\S}{\istensor{S}}
\renewcommand{\O}{\istensor{\Omega}}

\newcommand{\DelTen}{\istensor{\Delta}}
\newcommand{\GD}{\pseudovar{\G}}

\newcommand{\PGGt}{\P_{\GGt}}

\newcommand{\QG}{\Q_{\G}}
\newcommand{\QS}{\Q_{\S}}

\newcommand{\RG}{\R_{\G}}
\newcommand{\RS}{\R_{\S}}

\newcommand{\VTWO}{V^{2}}

\newcommand{\nutc}{\isvector{\nu}_{\subnut,c}} 
\newcommand{\nuts}{\isvector{\nu}_{\subnut,s}} 

\newcommand{\pnut}{\pseudovar{\nu}_{\subnut}} 
\newcommand{\pnutc}{\pseudovar{\isvector{\nu}}_{\subnut,c}} 
\newcommand{\pnuts}{\pseudovar{\isvector{\nu}}_{\subnut,s}} 

\newcommand{\trace}{\mathrm{tr}}
\newcommand{\tr}[1]{\trace(#1)}
\newcommand{\trpow}[2]{\trace^{#1}(#2)}
\newcommand{\traceless}[1]{{\tilde{#1}}}

\newcommand{\diag}{\mathrm{diag}}

\newcommand{\GGt}{\G\G\traspose}

\newcommand{\KH}{Kelvin-Helmholtz~}

\newcommand{\FLvol}{\Flength_{\mathrm{vol}}}
\newcommand{\FLSco}{\Flength_{\mathrm{Sco}}}
\newcommand{\FLmax}{\Flength_{\max}}
\newcommand{\FLmin}{\Flength_{\min}}
\newcommand{\FLLtwo}{\Flength_{\mathrm{L2}}}
\newcommand{\FLLapl}{\Flength_{\mathrm{Lapl}}}
\newcommand{\FLvort}{\Flength_{\vort}}
\newcommand{\FLMoc}{\tilde{\Flength}_{\vort}}
\newcommand{\FLSLA}{\Flength_{\mathrm{SLA}}}
\newcommand{\FLlsq}{\Flength_{\mathrm{lsq}}}
\newcommand{\FLrls}{\Flength_{\mathrm{rls}}}
\newcommand{\FLprls}{\tilde{\Flength}_{\mathrm{rls}}}
\newcommand{\Xavi}[1]{#1}
\newcommand{\XaviAlex}[1]{#1}
\newcommand{\XaviJesus}[1]{#1}

\newtheorem{remark}{Remark}

\newcommand{\befeq}[1]{}

\newcommand{\befmysubsection}[2]{}

\begin{document}


\title{A rational length scale for large-eddy simulation of turbulence on anisotropic grids}




\newcommand{\cttc}{\affiliation{Heat and Mass Transfer Technological Center, Technical University of Catalonia,\\
c/Colom 11, 08222 Terrassa, Spain}}

\newcommand{\ras}{\affiliation{Keldysh Institute of Applied Mathematics, 4A, Miusskaya Sq., Moscow 125047, Russia}}

\author{F.X.Trias}
\email[]{francesc.xavier.trias@upc.edu}
\cttc


\author{J.Ruano}
\email[]{jesus.ruano@upc.edu}
\cttc

\author{A.Duben}
\email[]{aduben@keldysh.ru}
\ras

\author{A.Gorobets}
\email[]{gorobets@keldysh.ru}
\ras


\date{\today}


\begin{abstract}
  \bef{Direct numerical simulations of the incompressible
    Navier--Stokes equations remain unfeasible for most real-world
    turbulent flows. Hence, dynamically less complex formulations are
    required for coarse-grained simulations.}  \aft{Due to the
    prohibitive cost of resolving all relevant scales, direct
    numerical simulations of turbulence remain unfeasible for most
    real-world applications. Consequently, dynamically simplified
    formulations are needed for coarse-grained simulations.} In this
  regard, eddy-viscosity models for Large-Eddy Simulation (LES) are
  widely used both in academia and industry. These models require a
  subgrid characteristic length, typically linked to the local grid
  size. While this length scale corresponds to the mesh step for
  isotropic grids, its definition for unstructured or anisotropic
  Cartesian meshes, such as the pancake-like meshes commonly used to
  capture near-wall turbulence or shear layers, remains an open
  question. Despite its significant influence on LES model
  performance, no consensus has been reached on its proper
  formulation. In this work, we introduce a novel subgrid
  characteristic length. This length scale is derived from the
  analysis of the entanglement between the numerical discretization
  and the filtering in LES. Its mathematical properties and simplicity
  \bef{suggest it is a robust choice that minimizes the impact of mesh
    anisotropies on simulation accuracy} \aft{make it a robust choice
    for reducing the impact of mesh anisotropies on simulation
    accuracy}. The effectiveness of the proposed subgrid length is
  demonstrated through simulations of decaying isotropic turbulence
  \aft{and a turbulent channel flow} \bef{on} \aft{using} different
  codes.
\end{abstract}



\maketitle 

\section{Introduction}

\label{intr}

\bef{The Navier--Stokes (NS) equations are an excellent mathematical
  model of turbulence. However, direct numerical simulations (DNS)}
\aft{Direct numerical simulations (DNS) of the Navier--Stokes (NS)
  equations} remain impractical for most real-world turbulent flows
because not enough resolution is available to resolve all the relevant
scales of motion. Therefore, practical numerical simulations have to
resort to turbulence modeling. Hence, we may turn to large-eddy
simulation (LES) to predict the large-scale behavior of turbulent
flows: namely, the large scales are explicitly computed, whereas
effects of small scale motions are modeled. Shortly, the LES equations
\XaviJesus{for incompressible flows} are obtained by applying a
spatial filter to the NS equations
\begin{equation}
\label{LESeq}
\partial_t \F{\vel} + ( \F{\vel} \cdot \grad ) \F{\vel} = \nu \lapl \F{\vel} - \nabla \F{p} - \nabla \cdot \tau \hspace{1mm}; \hspace{5mm} \nabla \cdot \F{\vel} = 0 ,
\end{equation}
\noindent where $\F{\vel}$ denotes the filtered velocity, and the
subgrid-scale (SGS) stress tensor \aft{is given by} $\tau = \F{\vel
  \otimes \vel} - \F{\vel} \otimes \F{\vel}$\aft{, where $\otimes$
  denotes the standard tensor product of vectors.} \bef{accounts for
  the effect of the unresolved scales.} The filter $\vel \rightarrow
\F{\vel}$, with filter length $\Flength$, is assumed to be symmetric
and to commute with differentiation. \bef{Since the} \aft{Notice that}
$\tau$ depends on both the filtered velocity, $\F{\vel}$, and the full
velocity field, $\vel$, \bef{this leads} \aft{leading} to a closure
problem. \bef{We thus have to approximate} \aft{Therefore,} $\tau$
\aft{is approximated} by a tensor \bef{depending only} \aft{that only
  depends} on the filtered velocity, \ie~$\tau \approx \tau ( \F{\vel}
)$.

\mbigskip

\bef{Due to its simplicity and robustness, the eddy-viscosity
  assumption remains by far the most widely used approach for closure
  modeling,} \aft{The eddy-viscosity assumption continues to be the
  dominant approach for closure modeling, primarily due to its
  simplicity and robustness,}
\begin{equation}
\label{eddyvis}
\tau ( \F{\vel} ) \approx - 2 \nut \S ( \F{\vel} ) ,
\end{equation}
\noindent where $\nut$ is the eddy viscosity, and $\S(\F{\vel}) =
\tfrac{1}{2} ( \grad \F{\vel} + \grad \F{\vel}\traspose )$ is the
rate-of-strain tensor. \bef{Note that the SGS stress tensor,
  $\tau(\F{\vel})$, is assumed to be traceless without loss of
  generality, since its trace can be included into the filtered
  pressure field, $\F{p}$. Then, most eddy-viscosity models can be
  formulated as follows:}\aft{The SGS tensor, $\tau(\F{\vel})$, is
  considered traceless, since its trace can be included into the
  filtered pressure term, $\F{p}$. With this simplification, most
  eddy-viscosity models take the following form:}
\begin{equation}
\label{eddyvis_template}
\nut = ( C_m \Flength )^2 D_m ( \F{\vel} ) ,
\end{equation}
\noindent where $C_m$ is the model constant, $\Flength$ denotes the
subgrid characteristic length, and $D_m(\F{\vel})$ is the
model-specific differential operator, with units of
frequency. \aft{The length scale $\Flength$, is the responsible for
  capturing the effective cut-off length scale, \ie~the spatial scale
  that separates the resolved turbulent motions, $\F{\vel}$, from the
  unresolved ones in an LES simulation. Then, the rest of the flow
  physics, such as the forward/backward scattering,
  laminar-to-turbulence transitions, 2D flow behavior or presence of
  walls must be captured by the the differential operator that defines
  the SGS model, \ie~$D_m ( \F{\vel} )$.}

\mbigskip

\bef{Over the past decades, LES research has mainly focused on either
  determining the model constant, $C_m$, or developing more accurate
  model operators $D_m(\F{\vel})$.} \aft{Over the past decades,
  advancements in LES have largely centered on calibrating the model
  constant $C_m$ and improving the accuracy of the model operator
  $D_m(\F{\vel})$.} One of the earliest milestones dates back to the
1960s, when Lilly~\cite{LIL67} proposed a method to compute the model
constant for the Smagorinsky model~\cite{SMA63}, showing excellent
performance in simulations of homogeneous isotropic
turbulence. However, the Smagorinsky model's differential operator,
$D_m ( \F{\vel} ) = | \S ( \F{\vel} ) |$, fails to vanish near walls,
leading to inaccurate results for wall-bounded flows. Initial
attempts to address this limitation relied on wall
functions~\cite{MOI82,PIO89}. A significant advancement in this regard
came with the dynamic procedure proposed by
Germano~\etal~\cite{GER91}, where the model constant $C_m$ is
determined using the Jacobi identity in a least-squares sense, as
originally proposed by Lilly~\cite{LIL92}. However, this method
results in highly fluctuating coefficient fields, often yielding
negative values for $\nut$, which can introduce numerical
instabilities. Consequently, additional techniques such as averaging
in homogeneous directions and {\it ad hoc} clipping of $\nut$ are
typically required. Due to these constraints, the original dynamic
procedure is not well-suited for flows in complex geometries that lack
homogeneous directions. To overcome these limitations, several
alternative approaches have been developed. Ghosal~\etal~\cite{GHO95}
introduced the dynamic localization model, while
Meneveau~\etal~\cite{MEN96} proposed the Lagrangian dynamic
model. Along similar lines, Park~\etal~\cite{PAR06} presented two
global dynamic models: one based on the Germano identity~\cite{GER92},
and another employing two test filters to enforce a “global
equilibrium” between viscous dissipation and subgrid-scale (SGS)
dissipation. Later, You and Moin~\cite{YOU07} proposed a dynamic
global approach utilizing only a single test filter. Tejada-Martínez
and Jansen~\cite{TEJ04,TEJ06} adopted an alternative strategy by
dynamically calculating the filter-width ratio, which is the only
model parameter in the dynamic Smagorinsky model.

\mbigskip

Alternatively, models that vanish near solid walls can be developed by
constructing appropriate differential operators, $D_m ( \F{\vel}
)$. The first examples thereof are the WALE~\cite{NIC99} and the
Vreman's model~\cite{VRE04b}. Later, Nicoud~\etal~\cite{NIC11}
introduced the $\sigma$-model, while Verstappen proposed the
$QR$-model \cite{VER11}, which was later generalized for anisotropic
grids by Rozema~\etal~\cite{ROZ15}, resulting in the anisotropic
minimum dissipation model. This list can be completed with the
eddy-viscosity model proposed by Ryu and Iaccarino~\cite{RYU14}, the
family of S3PQR models~\cite{TRI14-Rbased} proposed by co-authors of
this paper and the vortex-stretching-based eddy-viscosity
model~\cite{SIL17}. \XaviAlex{Finally, it is worth mentioning recent
  physics-constrained machine-learning SGS closures in which the
  near-wall scaling is imposed~\cite{HAS25}.}


\subsection{Computing the characteristic length scale: a short review}

Surprisingly, the computation of the subgrid characteristic length has
received comparatively little attention within the LES community, even
though it is a fundamental component of any eddy-viscosity model (see
Eq.\ref{eddyvis_template}). Moreover, due to its simplicity and
suitability for unstructured meshes, the most commonly adopted
definition is the one introduced by Deardorff\cite{DEA70}, which
defines $\Flength$ as the cube root of the cell volume. On a Cartesian
grid, it reads
\begin{equation}
\label{DeltaDeardorff}
\FLvol = ( \Dx \Dy \Dz )^{1/3} .
\end{equation}
Subsequently, Schumann~\cite{SCH75}, Lilly~\cite{LIL88}, and
Scotti~\etal~\cite{SCO93} proposed various adaptations of $\FLvol$ to
account for anisotropic grids. It was observed that the Deardorff's
length scale provides reasonable accuracy for small
anisotropies. However, \XaviAlex{corrections become necessary for
  highly anisotropic meshes, such as pancake-like meshes with
  significantly finer resolution in one spatial direction, which are
  typically used to capture near-wall turbulence or shear layers. In
  this regard,} Scotti~\etal~\cite{SCO93} proposed the following
correction:
\begin{equation}
\label{DeltaScotti}
\FLSco = f ( a_1 , a_2 ) \FLvol ,
\end{equation}
\noindent where $f ( a_1 , a_2 ) = \cosh \sqrt{ 4/27 [ ( \ln a_1 )^2 -
    \ln a_1 \ln a_2 + (\ln a_2)^2 ]}$ and $a_1 = \Dx/\Dz$, $a_2 =
\Dy/\Dz$, assuming that $\Dx \le \Dz$ and $\Dy \le \Dz$. \Xavi{Later
  on,} Scotti~\etal.~\cite{SCO97} tested the definition of
\Xavi{$\FLSco$} given in Eq.~(\ref{DeltaScotti}) for forced isotropic
turbulence using highly anisotropic grids. They compared the
\Xavi{theoretical} correction factor $f ( a_1, a_2 )$ with the
correction $f_{dyn} ( a_1, a_2)$ obtained by applying the dynamic
approach~\cite{GER91} to $C_m \FLvol f_{dyn} ( a_1, a_2 )$. \Xavi{To
  do so, they kept $C_m$ constant regardless of the mesh
  anisotropy}. \Xavi{Notice, that the dynamic approach is usually
  applied to find the model constant, $C_m$}. Their findings indicated
that the dynamic model captures the correct trend for pancake-like
grids ($a_2=1, \Dx \ll \Dy = \Dz$) but fails for pencil-like grids
($a_1=a_2, \Dx = \Dy \ll \Dz$). A further limitation of the approach
proposed by Scotti~\etal~\cite{SCO93} (see Eq.\ref{DeltaScotti}) is
that it is applicable only to structured Cartesian grids. To address
this, Colosqui and Oberai~\cite{COL08} developed an extension suitable
for unstructured meshes, assuming that the second-order structure
function adheres to the Kolmogorov's hypotheses. \XaviAlex{More recent
  works have focused on the construction~\cite{PRA24} and performance
  analysis of different SGS models~\cite{CHA23} and length
  scales~\cite{SCH20} on anisotropic grids.}

\mbigskip

Alternative definitions of the subgrid characteristic length scale,
$\Flength$, include the $L^2$-norm of the $3 \times 3$ diagonal tensor
\begin{equation}
\label{DelTen_def}
\DelTen \equiv \diag ( \Dx , \Dy , \Dz ) ,
\end{equation}
divided by $\sqrt{3}$,
\begin{equation}
\label{DeltaL2}
\FLLtwo = \sqrt{(\Dx^2 + \Dy^2 + \Dz^2)/3} ,
\end{equation}
\noindent and the square root of the harmonic mean of the squares of
the grid sizes
\begin{equation}
\label{DeltaLapl}
\FLLapl = \sqrt{3/(1/\Dx^2+1/\Dy^2+1/\Dz^2)} .
\end{equation}
\noindent \bef{which is directly related to the largest eigenvalue of
  the discrete approximation of minus the Laplacian operator,
  $-\lapl$. Furthermore, in the seminal work on the Detached-Eddy
  Simulation (DES) method, Spalart~\etal~\cite{SPA97} proposed using
  the maximum cell dimension,} \aft{This quantity is a good estimation
  of the largest eigenvalue (in absolute value) of the discrete
  Laplacian. Additionally, in their seminal work on the Detached-Eddy
  Simulation (DES) method, Spalart~\etal~\cite{SPA97} proposed using
  the maximum dimension of the cell,}
\begin{equation}
\label{Deltamax}
\FLmax = \max ( \Dx , \Dy , \Dz ) ,
\end{equation}
\noindent as a safer and robust definition of $\Flength$.

Over the past decade and a half, several subgrid characteristic length
scales have been proposed in the context of DES. \bef{Namely,
  Chauvet~\etal~\cite{CHA07} firstly introduced the concept of
  sensitizing $\Flength$ to the local flow topology.}
\aft{Chauvet~\etal~\cite{CHA07} were the first to propose making the
  subgrid length scale, $\Flength$, responsive to the local flow
  topology.} In particular, they proposed a new $\Flength$ that
depends on the orientation of the vorticity vector, ${\vort} = (
\vortx, \vorty, \vortz ) = \nabla \times {\vel}$,
\begin{equation}
\label{Deltavort}
\FLvort = \sqrt{ ( \vortx^2 \Dy \Dz + \vorty^2 \Dx \Dz + \vortz^2 \Dx \Dy ) / | \vort |^2} .
\end{equation}
\noindent The formulation was later extended to unstructured meshes by
Deck~\cite{DEC12}. This definition captures the alignment of the
vorticity vector, $\vort$, with the coordinate axis. For example, if
$\vort = ( 0 , 0 , \vortz )$, then $\FLvort$ simplifies to $\sqrt{\Dx
  \Dy}$. This approach was developed to address the issue of excessive
subgrid-scale dissipation introduced by $\FLmax$ in the early stages
of shear layers, which are commonly resolved using highly anisotropic
grids. In DES simulations, this excess dissipation artificially delays
the \KH instabilities by shifting the transition from RANS to LES mode
further downstream. Nevertheless, similar to $\FLvol$, the length
scale $\FLvort$ given in Eq.(\ref{Deltavort}) may still be influenced
by the smallest grid spacing, which can result in unrealistically low
eddy-viscosity values.

\mbigskip

To address this issue, Mockett~\etal~\cite{MOC15} proposed the
following flow-dependent length scale, 
\begin{equation}
\label{DeltaShu}
\FLMoc = \frac{1}{\sqrt{3}} \max_{n,m=1,\dots,8} | \isvector{l}_n - \isvector{l}_m | ,
\end{equation}
\noindent where $\isvector{l} = \vort/|\vort| \times \isvector{r}_n$
and $\isvector{r}_n$ ($n=1,\dots,8$ for a hexahedral cell) are the
locations of the cell vertices. The quantity $\FLMoc$ represents the
diameter of the set of cross-products points, $\isvector{l}_n$,
divided by $\sqrt{3}$. In the case with $\vort = ( 0, 0, \vortz )$, it
reduces to $\FLMoc = \sqrt{ ( \Dx^2 + \Dy^2 ) / 3 }$. Hence, it is
${\cal O} ( \max \{ \Dx, \Dy \} )$ instead of $\FLmax = \Dz$ (for the
common scenario where $\Dz > \Dx$ and $\Dz > \Dy$) or $\FLvort =
\sqrt{ \Dx \Dy }$. Thus, \textit{``unlike $\FLvort$ the
  definition~(\ref{DeltaShu}) never leads to a strong effect of the
  smallest grid-spacing on the subgrid-scale $\Flength$ even though it
  achieves the desired decrease compared to the $\FLmax$ definition in
  the quasi-2D flow regions treated on strongly anisotropic
  grids.''}~\cite{SHU15}.

\mbigskip

Later on, Shur~\etal~\cite{SHU15} proposed the Shear Layer Adapted
(SLA) subgrid length scale
\begin{equation}
\label{DeltaSLA}
\FLSLA = \FLMoc F_{KH} ( VTM ) .
\end{equation}
\noindent It is basically a modification of the $\FLMoc$ given in
Eq.(\ref{DeltaShu}), where the non-dimensional function $0 \le F_{KH}
( VTM ) \le 1$ is modulated by the Vortex Tilting Measure (VTM) given
by
\begin{equation}
VTM = \frac{| ( \S \cdot \vort ) \times \vort |}{\vort^2 \sqrt{-\Q_\traceless{\S}}} ,
\end{equation}
\noindent where $\traceless{\S}$ represents the traceless component of
the rate-of-strain tensor, $\S = 1/2 ( \nabla \F{\vel} + \nabla
\F{\vel}\traspose )$, \ie~$\traceless{\S} = \S - 1/3 \tr{\S}
\Identity$. For incompressible flows, where $\tr{\S} = \diver \F{\vel}
= 0$, it follows that $\traceless{\S} = \S$. \Xavi{Moreover,
  $\Q_\traceless{\S}$ denotes the second invariant of the tensor
  $\traceless{\S}$, given by $\Q_\traceless{\S} = 1/2 \left(
  \trpow{2}{\traceless{\S}} - \tr{\traceless{\S}^2} \right) = - 1/2
  \tr{\traceless{\S}^2}$}. The vortex tilting measure is bounded
within the range $0 \leq VTM \leq 1$ and equals zero when the
vorticity vector aligns with an eigenvector of $\S$ corresponding to
an eigenvalue $\lambda_i$, \ie~$\S \vort = \lambda_i \vort$. Thus, the
VTM quantifies the extent to which the rate-of-strain tensor tilts the
vorticity vector. Finally, the function $F_{KH}$ is designed to
trigger the \KH instability in the early stages of a \XaviAlex{shear
  layer~\cite{GUS17,XIA20}}, \XaviAlex{while in fully developed 3D
  turbulent regions, it should recover the original $\FLMoc$ length
  scale. In this regard,} Shur~\etal~\cite{SHU15} proposed several
forms for $F_{KH}$, all satisfying the fundamental constraints: $0
\leq F_{KH} ( VTM ) \leq 1$, with $F_{KH} ( 0 ) = 0$ and $F_{KH} ( 1 )
= 1$. \XaviAlex{Very recent studies evaluating the performance of
  $\FLSLA$ within the framework of DES models can be found in the
  literature~\cite{ZHO25,ZHA25}.}

\mbigskip

More recently, co-authors of the present paper proposed the $\FLlsq$
length scale~\cite{TRI16-CharLength}
\begin{equation}
\label{DeltaLsq}
\FLlsq = \sqrt{\frac{\GD \GD\traspose : \GGt}{\GGt : \GGt}} .
\end{equation}
\noindent where $\G \equiv \grad \F{\vel}$ is the gradient of the
resolved velocity and
\begin{equation}
\label{GD_def}
\GD \equiv \G \DelTen ,
\end{equation}
\noindent is the gradient in the computational space, \ie~without
considering the distances between adjacent nodes, and $\DelTen$ is the
second-order tensor containing the local mesh information given in
Eq.(\ref{DelTen_def}). The $\FLlsq$ length scale given in
Eq.(\ref{DeltaLsq}) follows from the least square minimization between
the tensors $\frac{\FLlsq^2}{12}\GGt$ and $\frac{1}{12}\GD
\GD\traspose$, respectively. These two tensors correspond to the Clark
model~\cite{CLA79},~\ie~the leading term in the Taylor-series
expansion of the SGS tensor $\tau$, assuming an isotropic filter
length equal to $\FLlsq$ and an anisotropic filter equal to $\DelTen =
\diag ( \Dx , \Dy , \Dz )$, respectively. This definition of
$\Flength$ has a list of desired properties (locality, boundedness,
flow dependence, ... ) for LES simulations. Moreover, it has been
shown to be a good approach as a grey area mitigation (GAM) technique
in the context of DES models \XaviAlex{particularly for
  aerodynamic~\cite{PONTRI20AIAA} and
  aeroacoustic~\cite{DUBRUATRI22AIAA} simulations. In the latter
  work~\cite{DUBRUATRI22AIAA}, an unstructured vertex-centered
  edge-based implementation of both $\FLvort$ and $\FLMoc$ was also
  presented. For further details on $\FLlsq$,} the reader is referred
to the original paper~\cite{TRI16-CharLength}.

\mbigskip

\bef{This work proposes a novel definition of the subgrid
  characteristic length, $\Flength$, aimed at addressing the following
  research question:} \aft{In this study, a new definition of the
  subgrid characteristic length, $\Flength$, is introduced, motivated
  by the following research question:} {\it can we establish a simple,
  robust, and easily implementable definition of $\Flength$ for any
  type of grid that minimizes the impact of mesh anisotropies on the
  performance of SGS models?} In doing so, we firstly analyse the
entanglement between the numerical discretization and the filtering in
LES. This is done in Section~\ref{FVM_filter} within a finite-volume
framework. Next, the new length scale, referred to as the {\it
  rational length scale}, is introduced in
Section~\ref{rls}. Additionally, implementation details are discussed,
along with an alternative dissipation-equivalent definition. This new
length scale is analysed and numerically tested in
Section~\ref{analysis}. Finally, relevant results are summarized and
conclusions are given.


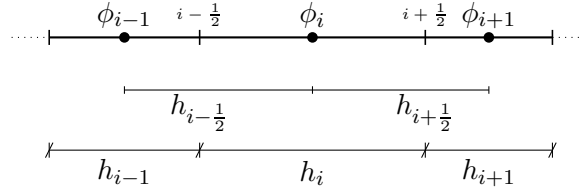
\begin{figure}[!t]
\centerline{
  \begin{tikzpicture}
    \def\ha{2}
    \def\hb{3}
    \def\hc{1.69}
    \def\hdots{0.5}
    \def\hticks{0.1}
    
    \draw[dotted] (-\hdots,0) -- (0, 0);
    \draw[dotted] (\ha+\hb+\hc,0) -- (\ha+\hb+\hc+\hdots,0);
    
    \draw[thick] (0,0) -- (\ha+\hb+\hc,0);
    
    \draw[semithick] (0,-\hticks) -- (0,\hticks);    
    \draw[semithick] (\ha,-\hticks) -- (\ha,\hticks);
    \draw[semithick] (\ha+\hb,-\hticks) -- (\ha+\hb,\hticks);
    \draw[semithick] (\ha+\hb+\hc,-\hticks) -- (\ha+\hb+\hc,\hticks);
    
    \node at (\ha,0.3) {\tiny $i-\frac{1}{2}$};
    \node at (\ha+\hb,0.3) {\tiny $i+\frac{1}{2}$};
    
    \fill (\ha/2,0) circle (2pt);
    \node at (\ha/2,0.3) {$\phi_{i-1}$};
    
    \fill (\ha+\hb/2.0,0) circle (2pt);
    \node at (\ha+\hb/2,0.3) {$\phi_{i}$};
    
    \fill (\ha+\hb+\hc/2,0) circle (2pt);
    \node at (\ha+\hb+\hc/2,0.3) {$\phi_{i+1}$};
    
    \draw[thin] (0,-1.5) -- (\ha+\hb+\hc,-1.5);
    \foreach \x in {0,\ha,\ha+\hb,\ha+\hb+\hc} {
        \draw[thin] (\x,-1.6) -- (\x,-1.4);
    }
    \foreach \x in {0,\ha,\ha+\hb,\ha+\hb+\hc} {
        \draw[thin] (\x-0.05,-1.6) -- (\x+0.05,-1.4);
    }
    
    \node at (\ha/2,-1.8) {$h_{i-1}$};
    \node at (\ha+\hb/2,-1.8) {$h_{i}$};
    \node at (\ha+\hb+\hc/2,-1.8) {$h_{i+1}$};

    \draw[thin] (\ha/2,-0.7) -- (\ha+\hb+\hc/2,-0.7);
    \foreach \x in {\ha/2,\ha+\hb/2,\ha+\hb+\hc/2} {
        \draw[thin] (\x,-0.75) -- (\x,-0.65);
    }
    \node at (\ha,-1) {\small $h_{i-\frac{1}{2}}$};
    \node at (\ha+\hb,-1) {\small $h_{i+\frac{1}{2}}$};
\end{tikzpicture}
}
\caption{Example of a one-dimensional mesh.}
\label{1Dmesh}
\end{figure}


\section{Finite-volume filtering}

\label{FVM_filter}

Let us consider a generic 1D convection-diffusion equation
\begin{equation}
\label{conv-diff_eq}
\frac{\partial \sca}{\partial t} + \frac{\partial ( \uvel \sca )}{\partial x} = \frac{\partial}{\partial x} \left( \Gamma \frac{\partial \sca}{\partial x} \right) ,
\end{equation}
\noindent where $\uvel(x,t)$ denotes the advective velocity,
$\sca(x,t)$ represents a generic (transported) scalar field \Xavi{and
  $\Gamma(x)$ is the diffusivity}. Within a finite-volume method
(FVM), this equation is integrated over a set of non-overlapping
volumes. Hence, discrete FVM variables result by applying a box filter
with filter width equal to the local grid size, $h$,
\begin{equation}
\Filter{\sca} (x) = \frac{1}{h} \int_{x-\halving{h}}^{x+\halving{h}} \phi \ud x .
\end{equation}
\noindent Notice that this filter commutes with differentiation
\begin{equation}
\label{box_commutation}
\frac{\partial \Filter{\sca}}{\partial x} = \frac{1}{h} \frac{\partial}{\partial x} \int_{x-\halving{h}}^{x+\halving{h}} \sca \ud x = \frac{1}{h} \int_{x-\halving{h}}^{x+\halving{h}} \frac{\partial \sca}{\partial x} \ud x = \Filter{\frac{\partial \sca}{\partial x}} .
\end{equation}
\noindent Moreover, the standard second-order approximation of the
first-derivative at the face is exactly equal to the filtered
derivative
\begin{equation}
\label{FilteredGrad}
\left. \frac{\partial \sca}{\partial x} \right|_{i+\half} \approx \frac{\sca_{i+1} - \sca_{i}}{h_{i+\half}} = \frac{1}{h_{i+\half}} \int_{x_i}^{x_{i+1}} \frac{\partial \sca}{\partial x} \ud x = \left. \Filter{\frac{\partial \sca}{\partial x}} \right|_{i+\half} \stackrel{Eq.(\ref{box_commutation})}{=} \left. \frac{\partial \Filter{\sca}}{\partial x} \right|_{i+\half} .
\end{equation}
\noindent This example corresponds to the 1D non-uniform mesh displayed in Figure~\ref{1Dmesh}.
\begin{remark}
This result suggests that the actual filter length when computing the
face derivative is $h_{i+\half}$, \ie~the distance between the
adjacent nodes $i$ and $i+1$ (see Figure~\ref{1Dmesh}).
\end{remark}

\mbigskip

Finally, the diffusive term in a FVM framework is approximated as follows
\begin{align}
\nonumber \left. \frac{\partial}{\partial x} \left( \Gamma \frac{\partial \sca}{\partial x} \right) \right|_{i} &\approx 
\frac{1}{h_i} \left( \left. \Gamma \frac{\partial \sca}{\partial x} \right|_{{i+\frac{1}{2}}} - \left.\Gamma \frac{\partial \sca}{\partial x} \right|_{{i-\frac{1}{2}}}\right) =
\left. \Filter{\frac{\partial}{\partial x} \left( \Gamma \frac{\partial \sca}{\partial x} \right)} \right|_{i} \\
\label{BoxDiff} &\approx \frac{1}{h_i} \left( \Gamma_{i+\half} \frac{\sca_{i+1} - \sca_{i}}{h_{i+\half}} - \Gamma_{i-\half} \frac{\sca_{i} - \sca_{i-1}}{h_{i-\half}} \right) =
\left. \Filter{\frac{\partial}{\partial x} \left( \Gamma \Filter{\frac{\partial \sca}{\partial x}} \right)} \right|_{i} .
\end{align}
In the case of non-constant diffusivity, $\Gamma(x)$, its evaluation
at the cell faces can be viewed as box filtering the $\Gamma$
field,~\ie
\begin{equation}
\label{FilteredGamma}
\Gamma_{i+\half} \approx \frac{\Gamma_{i} + \Gamma_{i+1}}{2} \approx \frac{1}{h_{i+\half}} \int_{x_{i}}^{x_{i+1}} \Gamma(x) \ud x = \Filter{\Gamma}_{i+\half} ,
\end{equation}
\noindent while using the trapezoidal rule to approximate the
integral. Altogether leads to
\begin{align}
\label{FVMfilter}
\nonumber \left. \frac{\partial}{\partial x} \left( \Gamma \frac{\partial \sca}{\partial x} \right) \right|_{i} &\approx 
\frac{1}{h_i} \left( \frac{\Gamma_{i+1} + \Gamma_i}{2} \frac{\sca_{i+1} - \sca_{i}}{h_{i+\half}} - \frac{\Gamma_{i} + \Gamma_{i-1}}{2} \frac{\sca_{i} - \sca_{i-1}}{h_{i-\half}} \right) \\
&= \left. \FilterFVM{\frac{\partial}{\partial x} \left( \Filter{\Gamma} \Filter{\frac{\partial \sca}{\partial x}} \right)} \right|_{i} ,
\end{align}
\noindent where $\FilterFVM{(\cdot)}$ is the FVM filtering along the
cell $i$ whereas the other two filters are applied to the face
quantities and their associated filter lengths are $h_{i+\half}$ and
$h_{i-\half}$, respectively.

\begin{remark}
This result suggests that two filtering operations are performed when
computing the diffusive term: the calculation of the face derivative
(see Eq.\ref{FilteredGrad}) and the cell-to-face interpolation of the
diffusivity coefficient (see Eq.\ref{FilteredGamma}). Both filtering
operators share the same filter length; namely, the distance between
the nodes adjacent to the corresponding face.
\end{remark}

\noindent Consequently, this suggests that the SGS characteristic
length used to compute the eddy viscosity, $\nut$, at the face
$i+\half$ should be $h_{i+\half}$ (see Figure~\ref{1Dmesh}). This
concept forms the foundation of this paper and is extended to general
meshes in the following section.


\begin{figure}[!t]
  \centering{ \includegraphics[height=0.46\textwidth,angle=0]{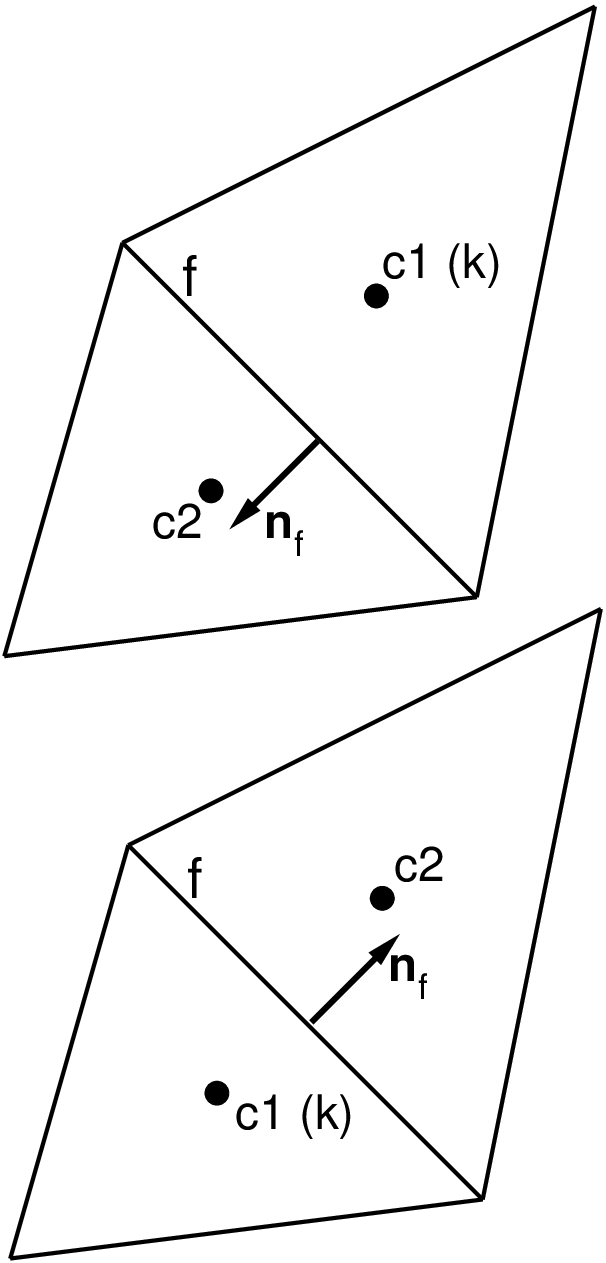} \hspace{9.69mm} \includegraphics[height=0.43\textwidth,angle=0]{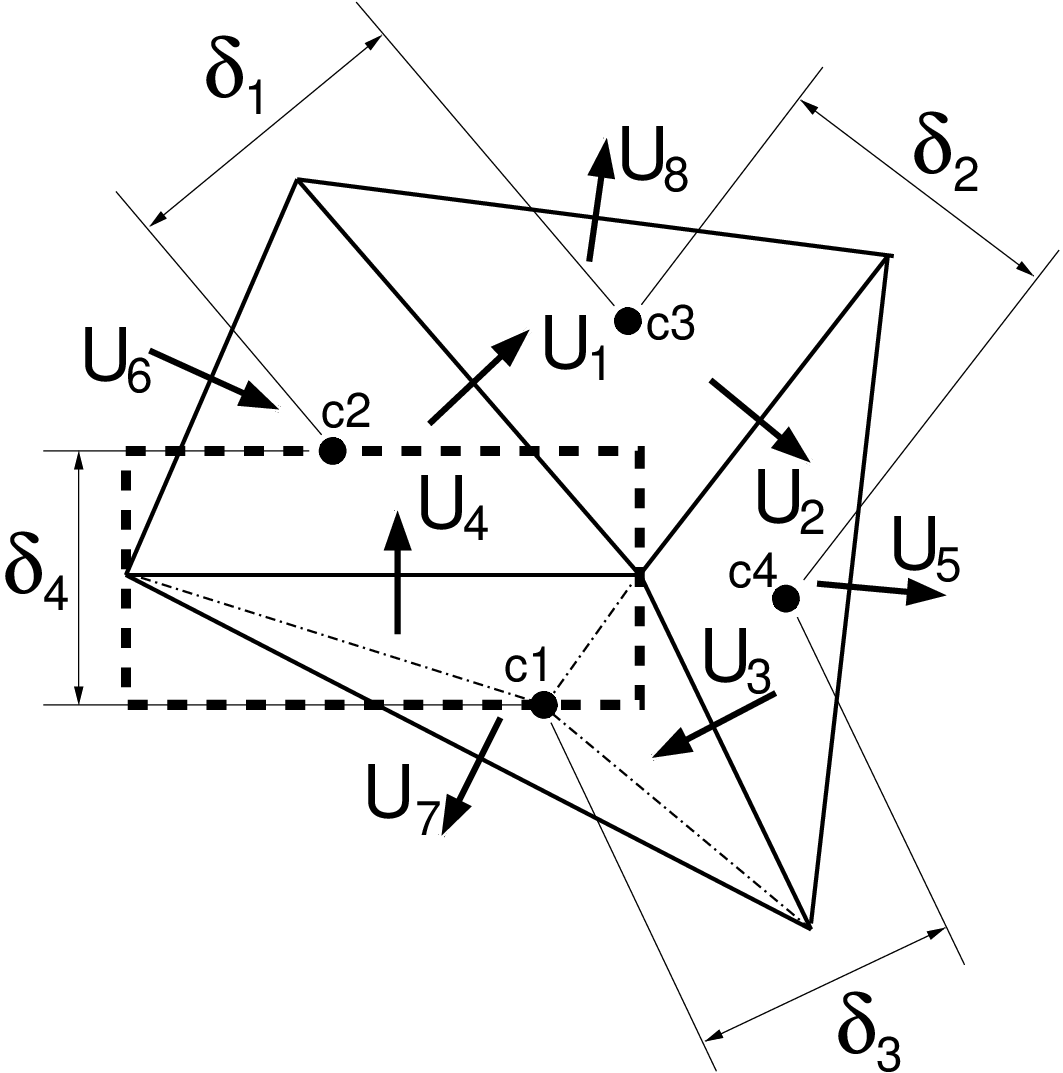}
    }
\caption{Left: face normal and neighbor labeling criterion. Right:
  definition of the volumes, $\vcvects$, associated with the the
  face-normal velocities, $\velh$. Thick dashed rectangle is the
  volume associated with the staggered velocity $\mathrm{U}_{4} = [
    \velh ]_{4}$, \ie~$[ \vcvects ]_{4,4} = A_{4} \delta_{4}$ where
  $A_{4}$ is the face area and $\delta_{4} = | \normal_{4} \cdot
  \protect \overrightarrow{\cpA \cpB} |$ is the projected distance
  between adjacent cell centers. Thin dash-dotted lines are placed to
  illustrate that the sum of volumes is exactly preserved
  $\trace{\vcvects}=\trace{\vcvect}= d \trace{\vcvectc}$ ($d=2$ for 2D
  and $d=3$ for 3D) regardless of the location of the cell nodes.}
\label{mesh}
\end{figure}


\section{A new length scale approach}

\label{rls}

\subsection{Rational length scale}

The formula given in Eq.(\ref{FVMfilter}) can be generalized for
multiple dimensions (also unstructured grids). Namely, let us first
consider the FVM discretization of the diffusive term on an arbitrary
mesh
\begin{equation}
\label{grad3D}
\diver ( \Gamma \grad \sca ) \approx \vcvectc^{-1} \dive \diffm \gradd \scafieldc .
\end{equation}
\XaviAlex{Unless otherwise stated,} we follow the FVM notation used
in previous works~\cite{TRI08-JCP,TRI22-AlgEigCD}. Namely, in
Eq.(\ref{grad3D}), $\gradd$ denotes the cell-to-face discrete gradient
operator,
\begin{equation}
\label{gradd}
\gradd \equiv \DX^{-1} \Tcs ,  
\end{equation}
\noindent where $\Tcs$ is the cell-to-face oriented incidence
matrix. \XaviAlex{The sub-index $h$ is used here to stress that
  $\gradd$ is a discrete operator and to avoid confusion with the
  gradient tensor of the resolved velocity, $\G \equiv \grad
  \F{\vel}$. Notice that $\Tcs$} has two non-zero elements per row (a
$+1$ and a $-1$ corresponding to the cells adjacent to a face),
whereas the face-to-cell incidence matrix, $\Tsc$, corresponds to its
transpose, $\Tsc = \Tcs\traspose$. For example, for the mesh with
$4$~control volumes and $8$~faces shown in Figure~\ref{mesh} (right),
the latter reads
\begin{equation}
\label{Tsc_def}
\Tsc = \Tcs\traspose =
\left(
\begin{array}{rrrrrrrr}
 0 &  0 & -1 & +1 &  0 &  0 & +1 &  0 \\
+1 &  0 &  0 & -1 &  0 & -1 &  0 &  0 \\
-1 & +1 &  0 &  0 &  0 &  0 &  0 & +1 \\
 0 & -1 & +1 &  0 & +1 &  0 &  0 &  0
\end{array}
\right) .
\end{equation}
\noindent Moreover, the diagonal matrix $\DX$ contains the projected
distances between adjacent cell \Xavi{centers} (see~$\delta_i$ with
$i=\{1,\dots,4\}$ in Figure~\ref{mesh}). Therefore, the formula in
Eq.(\ref{grad3D}) corresponds to the standard second-order gradient
calculation, which, in 1D, leads to the result given in
Eq.(\ref{FilteredGrad}). Notice that subindices $c$ and $s$ refers to
whether quantities are located at the cell nodes or staggered at the
faces.

\mbigskip

Then, the integrated divergence operator is defined as follows
\begin{equation}
\label{dive}
\dive \equiv - \gradd\traspose \vcvects \stackrel{(\ref{gradd})}{=} - \Tcs\traspose \DX^{-1} \vcvects = - \Xavi{\Tcs\traspose} \As .
\end{equation}
In this way, the duality between the gradient and the divergence
operators is preserved at discrete level. Notice that $\vcvects \equiv
\DX \As$ is a diagonal matrix containing the staggered volumes whereas
$\As$ is a diagonal matrix containing the corresponding surface
areas. Finally, $\vcvectc$ and $\diffm$ are diagonal matrices
containing the cell volumes and the diffusivities associated with the
faces, respectively.

\mbigskip

Then, combining Eqs.(\ref{grad3D}), (\ref{gradd}) and (\ref{dive}), we
obtain
\begin{equation}
\label{FVMdiff1}
\diver ( \Gamma \grad \sca ) \approx - \vcvectc^{-1} \underbrace{\Tcs\traspose \As}_{\XaviJesus{-\dive}} \diffm \underbrace{\DX^{-1} \Tcs}_{\XaviJesus{\gradd}} \scafieldc = - \vcvectc^{-1} \underbrace{\Tcs\traspose \DX^{-1} \vcvects}_{\XaviJesus{-\dive}} \diffm \underbrace{\DX^{-1} \Tcs}_{\XaviJesus{\gradd}} \scafieldc .
\end{equation}
Here, we can neatly identify the filter lengths associated with the
box filtering applied to the gradient (see Eq.~\ref{FilteredGrad});
that is, the matrix $\DX$ that contains the projected distances
between the adjacent cell \Xavi{centers}.

\mbigskip

Hence, if diffusivities $\diffm$ are replaced by $\pdiffm$, which is a
diagonal matrix containing the inverses of the characteristic
relaxation times, multiplied by their associated length scales, it
reads
\begin{equation}
\label{diffm_def}
\diffm \equiv \DX \pdiffm \DX .
\end{equation}
Plugging this into Eq.(\ref{FVMdiff1}) leads to
\begin{equation}
\label{FVMdiff2}
\dive ( \Gamma \Xavi{\gradd} \sca ) \approx - \vcvectc^{-1} \Tcs\traspose \XaviJesus{\vcvects} \pdiffm \Tcs \scafieldc .
\end{equation}
Hence, we are computing the gradient (and its transpose) without
distances, \ie~in the so-called computational space.

\mbigskip

An alternative interpretation is the following: going back to the
original formula given in Eq.(\ref{grad3D})
\begin{equation}
\label{DeltaRLS}
\diver ( \Gamma \grad \sca ) \approx \vcvectc^{-1} \dive \diffm \gradd \scafieldc \stackrel{(\ref{diffm_def})}{=} \vcvectc^{-1} \dive \DX^2 \pdiffm \gradd \scafieldc .
\end{equation}
Here, it is very easy to identify the analogy with the formula given
in Eq.(\ref{FVMfilter}). This naturally leads to a definition of the
length scale for SGS models, $\DX$, which is computed directly at the
faces and corresponds to the distance between adjacent cell centers
(see~$\delta_i$ with $i=\{1,\dots,4\}$ in
Figure~\ref{mesh}). Hereafter, this new length scale is denoted
by~$\FLrls$ and referred to as {\it rational length scale}, as it
naturally arises from the entanglement between numerical
discretization and the LES filtering.

\begin{figure}[!t]
\centering
\begin{subfigure}[b]{0.99\textwidth}
\begin{tikzcd}
\velhc \arrow[r] & \graddc \velhc \arrow[r] & \{ \QG, \QS, \RG, \RS, \VTWO \} \arrow[r,"{\shortstack{\tiny {\it SGS} \\ \tiny {\it model}}}"] & \nutc \arrow[r, "\OIcs"] & \nuts & \\
  & & & \diffmc \arrow[r] \arrow[u, "\diag", sloped, leftrightarrow ] & \diffm \arrow[u, "\diag", sloped, leftrightarrow ]
\end{tikzcd}
\caption{\XaviAlex{Standard flowchart to computed the eddy-viscosity at the cell faces, $\nuts$.}}
\label{flowchart_standard}
\end{subfigure}
\hfill
\begin{subfigure}[b]{0.99\textwidth}
\begin{tikzcd}
\velhc \arrow[r] & \graddc \velhc \arrow[r] & \{ \QG, \QS, \RG, \RS, \VTWO \} \arrow[r,"{\shortstack{\tiny {\it SGS} \\ \tiny {\it model}}}"] & \pnutc \arrow[r, "\OIcs"] & \pnuts \arrow[r, "\DX^2"] & \nuts \\
  & & & \pdiffmc \arrow[r] \arrow[u, "\diag", sloped, leftrightarrow ] & \pdiffm \arrow[u, "\diag", sloped, leftrightarrow ]
\end{tikzcd}
\caption{\XaviAlex{Modified flowchart to computed the eddy-viscosity at the cell faces, $\nuts$.}}
\label{flowchart_new}
\end{subfigure}
\hfill
\begin{subfigure}[b]{0.99\textwidth}
\begin{tikzcd}
\nutc \arrow[r, "\vcvectc^{-2/3}", dashed] & \pnutc \arrow[r, "\OIcs"] & \pnuts \arrow[r, "\DX^2", dashed] & \nuts \\
\diffmc \arrow[r, dashed] \arrow[u, "\diag", sloped, leftrightarrow] & \pdiffmc \arrow[r] \arrow[u, "\diag", sloped, leftrightarrow ] & \pdiffm \arrow[r, dashed] \arrow[u, "\diag", sloped, leftrightarrow ] & \diffm \arrow[u, "\diag", sloped, leftrightarrow]
\end{tikzcd}
\caption{\XaviAlex{Flowchart for the implementation of the
    $\FLrls$. Dashed lines represent the modifications required
    respect to the standard flowchart.}}
\label{flowchart_implementation}
\end{subfigure}
\caption{\XaviAlex{Flowcharts to compute the eddy-viscosity at the cell faces, $\nuts$.}}
\label{flowcharts}
\end{figure}
%


\subsection{Practical implementation and an alternative definition}

\label{implementation}

The newly proposed length scale, $\FLrls$, is directly computed at the
faces as the distance between adjacent cell centers. However, in
virtually all codes, the turbulent viscosity is computed at cell
centers and then interpolated to the faces. The overall process is
depicted in the flowchart displayed in
Figure~\ref{flowchart_standard}. Namely, firstly the gradient of the
resolved velocity field, $\graddc \velhc$ is computed at the cells;
then, the eddy-viscosity, $\nutc$, is calculated at the cells based on
the five basic invariants of the velocity gradient tensor, as
virtually all existing models can be reformulated as combinations of
these invariants~\cite{TRI14-Rbased}. Then, $\nutc$ is interpolated to
the faces, \ie~$\nuts = \OIcs \nutc$, where $\OIcs$ is a cell-to-face
interpolation defined as
\begin{equation}
\label{OIcs}
\OIcs \equiv \frac{1}{2} | \Tcs | ,
\end{equation}
\noindent which is the cell-to-face midpoint interpolation. In our
case, we need to slightly modify the \Xavi{flowchart} as shown in
Figure~\ref{flowchart_new}. Specifically, the only required
modifications are indicated by dashed lines in the flowchart displayed
in Figure~\ref{flowchart_implementation}. Thus, the two changes are
basically two geometrical re-scalings. Firstly, if we assume that
$\FLvol$ is used in the calculation of $\nutc$, we need to divide it
by $\vcvectc^{2/3}$, which is the diagonal matrix that contains the
cell volumes, \ie~$\FLvol$ at cell $i$ is given
by~$[\vcvectc^{1/3}]_{i,i}$. In this way, we obtain a
pseudo-eddy-viscosity, $\pnutc$, computed without any length scale and
thus having units of inverse time. Then, a second re-scaling is
required to compute the actual eddy-viscosity at the faces, $\nuts$,
from $\pnuts$, \ie~$\nuts \equiv \DX^{2} \pnuts$.

\mbigskip

Therefore, from an implementation and conceptual point-of-view, the
new approach $\FLrls$ is completely different from all previous
definitions of $\Flength$ since it is directly computed at the
faces. Although the required code modifications are minimal, it may be
of interest to compute a length scale at the cells that provides the
same dissipation. Namely, it can be shown that the local dissipation
of the viscous term, with constant viscosity, is given by
\begin{equation}
\nu \G:\G = \nu \trace ( \G \G \traspose ) .
\end{equation}
If we replace $\nu$ by $\nut$, we can obtain a very accurate
estimation of the local dissipation introduced by an eddy-viscosity
model. Furthermore, in the new approach we also need to replace $\nut$
and $\G$ by $\pnut$ and $\GD$ (see Eq.\ref{GD_def}), respectively,
leading to
\begin{equation}
\pnut \GD:\GD = \pnut \trace ( \GD \GD\traspose ) .
\end{equation}
Then, we can compute an equivalent filter length, $\FLprls$, that
leads to the same local dissipation,~\ie
\begin{equation}
\label{DeltaPRLS}
\FLprls^2 \pnut \G:\G = \pnut \GD:\GD \hphantom{kkk} \Longrightarrow \hphantom{kkk} \boxed{\FLprls = \sqrt{\frac{\GD:\GD}{\G:\G}} = \sqrt{\frac{\trace(\GD \GD\traspose)}{\trace(\GGt)}}}
\end{equation}
Notice that $\PGGt = \trace ( \GGt )$ is the first invariant of the
symmetric tensor $\GGt$.


\begin{table}
\begin{center}
\aft{
\begin{tabular}{c|c|c}
Property & Relevance & Description \\
\hline
{\bf P1} & must      & positive ($\Flength > 0$), local and frame invariant \\
{\bf P2} & must      & bounded, \eg~for Cartesian mesh with $\Dx \le \Dy \le \Dz$, $\Dx \le \Flength \le \Dz$ \\
{\bf P3} & optional  & sensitive to the local flow field, $\G$ \\
{\bf P4} & practical & applicable to unstructured meshes \\
{\bf P5} & practical & length scale is directly computed at the cell faces \\
{\bf P6} & efficient & computational cost \\
{\bf P7} & efficient & memory footprint
\end{tabular}
}
\end{center}
\caption{\aft{List of potential properties of the subgrid characteristic
    length, $\Flength$.}}
\label{list_of_properties}
\end{table}

\newcommand{\mySize}{\tiny}
\newcommand{\myVeryLow}{\mySize $+$}
\newcommand{\myLow}{\mySize $++$}
\newcommand{\myMedium}{\mySize $+++$}
\newcommand{\myHigh}{\mySize $++++$}

\begin{table}
\begin{center}
\begin{tabular}{c|ccccccccc|cc}
         & $\FLvol$                  & $\FLSco$               & $\FLmax$            & $\FLLtwo$          & $\FLLapl$            & $\FLvort$            & $\FLMoc$            & $\FLSLA$            & $\FLlsq$            & $\FLrls$            & $\FLprls$     \\
\hline
Formula  & Eq.(\ref{DeltaDeardorff}) & Eq.(\ref{DeltaScotti}) & Eq.(\ref{Deltamax}) & Eq.(\ref{DeltaL2}) & Eq.(\ref{DeltaLapl}) & Eq.(\ref{Deltavort}) & Eq.(\ref{DeltaShu}) & Eq.(\ref{DeltaSLA}) & Eq.(\ref{DeltaLsq}) & Eq.(\ref{DeltaRLS}) & Eq.(\ref{DeltaPRLS}) \\
\hline
{\bf P1} & Yes                       & Yes                    & Yes                 & Yes                & Yes                  & Yes                  & Yes                 & Yes                 & Yes                 & Yes                 & Yes \\
{\bf P2} & Yes                       & Yes                    & Yes                 & Yes                & Yes                  & Yes                  & Yes                 & No                  & Yes                 & Yes                 & Yes \\
{\bf P3} & No                        & No                     & No                  & No                 & No                   & Yes                  & Yes                 & Yes                 & Yes                 & No$^{c}$                 & Yes \\
{\bf P4} & Yes                       & No                     & No$^{a}$             & No                 & No$^{a}$             & No$^{b}$              & Yes                 & Yes                 & Yes                 & Yes                 & Yes \\
{\bf P5} & No                        & No                     & No                  & No                 & No                   & No                   & No                  & No                  & No                  & Yes                 & No \\
{\bf \XaviAlex{P6}} & \myVeryLow                & \myVeryLow             & \myVeryLow          & \myVeryLow         & \myVeryLow           & \myLow               & \myMedium            & \myHigh             & \myMedium          & \myVeryLow          & \myMedium \\
{\bf \XaviAlex{P7}} & \myVeryLow                & \myVeryLow             & \myVeryLow          & \myVeryLow         & \myVeryLow           & \myLow               & \myLow            & \myMedium             & \myMedium          & \myVeryLow          & \myMedium
\end{tabular}
\end{center}
\caption{Properties of different definitions of the subgrid
  characteristic length, $\Flength$. \bef{Namely, {\bf P1}: positive
    ($\Flength \ge 0$), local and frame invariant; {\bf P2}: bounded,
    \ie~given a structured Cartesian mesh where $\Dx \le \Dy \le \Dz$,
    $\Dx \le \Flength \le \Dz$; {\bf P3}: sensitive to the local flow
    field; {\bf P4}: applicable to unstructured meshes; {\bf P5}:
    length scale is directly computed at the cell faces; {\bf P6}:
    computational cost\Xavi{; {\bf P7}: memory footprint}.} \aft{See
    Table~\ref{list_of_properties} for a short description of these
    properties.} $^{a}$ Possible with some adaptations, $^{b}$
  Deck~\cite{DEC12} proposed a generalization for unstructured
  meshes. $^{c}$ $\FLrls$ is computed at the faces independently of
  the local flow field, however, its effect ultimately depends on it.}
\label{properties_Delta}
\end{table}


\section{Analysis of the new approach and numerical tests}

\label{analysis}

\subsection{Fundamental properties}

\label{properties}

Starting with the classical Smagorinsky model~\cite{SMA63}, numerous
eddy-viscosity models (see Eq.\ref{eddyvis}) have been developed over
the past decades~\cite{SAG05}. The definition of $\nut$ in
Eq.(\ref{eddyvis_template}) serves as a general framework for most of
these models, requiring a subgrid characteristic length, $\Flength$,
which is typically related to the local grid size. In isotropic grids,
$\Flength$ corresponds to the mesh size,
\ie~$\Flength=\Dx=\Dy=\Dz$. However, there is still no consensus on
how to define $\Flength$ for (highly) anisotropic or unstructured
grids. Despite its potential inaccuracies in certain cases, more than
four decades later, the approach introduced by Deardorff~\cite{DEA70},
\ie~using the cube root of the cell volume (see
Eq.\ref{DeltaDeardorff}), remains the most widely used method in both
academia and industry.

\mbigskip

\bef{Alternative methods for computing $\Flength$ have been reviewed
  in Section~\ref{intr}. They are categorized in
  Table~\ref{properties_Delta} based on a set of desirable properties
  derived from physical, numerical, and practical considerations. The
  first property, denoted as {\bf P1}, ensures both positiveness and
  locality. While negative values of $\nut$ may be justified from a
  physical perspective due to the backscatter phenomenon, numerically,
  the condition $\nut \geq 0$ is generally preferred to guarantee the
  stability of the simulation.} \aft{Section~\ref{intr} provides an
  overview of alternative approaches for computing $\Flength$, which
  are classified in Table~\ref{properties_Delta} according to a set of
  desirable criteria (see Table~\ref{list_of_properties}) grounded in
  physical reasoning, numerical stability, and practical
  implementation. The first criterion, labeled as {\bf P1}, requires
  the length scale to be both positive and local. Although negative
  values of $\nut$ may be justified from a physical perspective due to
  the backscatter phenomenon, from a numerical standpoint, enforcing
  $\nut \geq 0$ is usually preferred to ensure simulation stability.}
Additionally, to preserve Galilean invariance, a fundamental physical
principle, flow-dependent definitions of $\Flength$ (see property {\bf
  P3} below) are formulated using invariants of the velocity gradient,
$\G \equiv \nabla \F{\vel}$. This approach guarantees locality, which
is particularly beneficial for applications involving complex
flows. The second property [{\bf P2}] ensures that $\Flength$ remains
properly bounded. Specifically, for a structured Cartesian mesh where
$\Dx \leq \Dy \leq \Dz$, we require that $\Dx \leq \Flength \leq
\Dz$. All length scales presented in Table~\ref{properties_Delta}
satisfy properties {\bf P1} and {\bf P2}, except the $\FLSLA$ that can
eventually lead to values of $\Flength$ smaller than the smallest grid
size. The third property [{\bf P3}] classifies the approaches for
computing $\Flength$ into two categories: \bef{those that depend
  solely on the geometrical properties of the mesh and those that also
  incorporate local flow topology, specifically the velocity gradient,
  $\G$.} \aft{into purely geometry-based definitions and those that
  incorporate flow-dependent information, particularly the velocity
  gradient, $\G$.}

Examples of the former category have been introduced in the
context of DES by Chauvet~\etal~\cite{CHA07} (see the definition of
$\FLvort$ in Eq.\ref{Deltavort}), Mockett~\etal~\cite{MOC15} (see
$\FLMoc$ in Eq.\ref{DeltaShu}), the modified version $\FLSLA$ proposed
by Shur\etal~\cite{SHU15} (see Eq.\ref{DeltaSLA}) and the definition
of $\FLlsq$ proposed by co-authors of this
paper~\cite{TRI16-CharLength} (see Eq.\ref{DeltaLsq}). This list is
further expanded with $\FLprls$, given in Eq.(\ref{DeltaPRLS}). Notice
that $\FLrls$, given in Eq.(\ref{DeltaRLS}), is computed at the faces
independently of the local flow field, however, its effect ultimately
depends on it, as remarked in Table~\ref{properties_Delta}.

\mbigskip

The \Xavi{last four} properties of the subgrid characteristic length
scale, $\Flength$, are mainly of practical significance. Property {\bf
  P4} refers to its applicability to unstructured meshes. In this
regard, all flow-dependent definitions of $\Flength$, satisfy this
property, including the newly proposed $\FLprls$ length
scale. However, among the definitions of $\Flength$ that are
independent of the local flow, only the Deardorff’s
approach~\cite{DEA70} and the newly proposed length scale $\FLrls$ can
be directly applied to unstructured grids. Additionally, $\FLrls$ is
the only length scale computed directly at the faces, as stated by
property {\bf P5}. All other length scales are first computed at the
cell and then interpolated to the faces. This interpolation can, in
fact, be seen as an unnecessary cell-to-face (box) filtering operation
(see Eq.\ref{FilteredGamma}). Finally, it is also important that the
definition of $\Flength$ is well-conditioned and computationally
efficient, as stated in property {\bf P6}. \Xavi{Moreover, a low
  memory footprint (property {\bf P7}) is another key feature that
  makes LES simulations more affordable on modern HPC systems.}
However, flow-dependent formulations can pose challenges in
\Xavi{these last two properties}, often demanding significantly higher
computational resources. Moreover, they must be carefully implemented
to avoid numerical issues, particularly when dealing with
indeterminate forms such as $0/0$. In this regard, the newly proposed
length scale $\FLrls$ is virtually costless, \Xavi{both in terms of
  computational cost and memory,} as it only needs to be computed once
at the faces, independently of the local flow field.


\subsection{Analysis of simple flows}

To get a better understanding of $\FLrls$ and $\FLprls$, we firstly
analyze several special cases. Firstly, for isotropic meshes they
reduce to $\Flength=\Dx=\Dy=\Dz$ regardless of the flow topology,
whereas for anisotropic grids they always remain well-bounded
(property {\bf P2}). Secondly, for pure rotation flows,
\ie~$\S=\istensor{0}$ and $\G = \O$, $\FLprls$ leads to
\begin{equation}
\label{Drls_rot}
\FLprls = \frac{\vortx^2 ( \Dy + \Dz ) + \vorty^2 ( \Dx + \Dz ) + \vortz^2 ( \Dx + \Dy )}{ 2 | \vort |^2} ,
\end{equation}
\noindent which resembles the definition of $\FLvort$ proposed by
Chauvet~\etal~\cite{CHA07} given in Eq.~(\ref{Deltavort}). Actually,
similarly to the definition of $\FLMoc$ proposed by
Mockett~\etal~\cite{MOC15} given in Eq.~(\ref{DeltaShu}), $\FLprls$
(also $\FLrls$) is ${\cal O} ( \max \{ \Dx, \Dy \} )$ instead of
$\FLvort = \sqrt{ \Dx \Dy }$ in the case where $\vort = ( 0 , 0 ,
\vortz )$. Therefore, it also avoids strong effects of the smallest
grid-spacing in the initial part of a shear layer.
\begin{figure}[!t]
\centering{
\includegraphics[height=0.6969\textwidth,angle=-90]{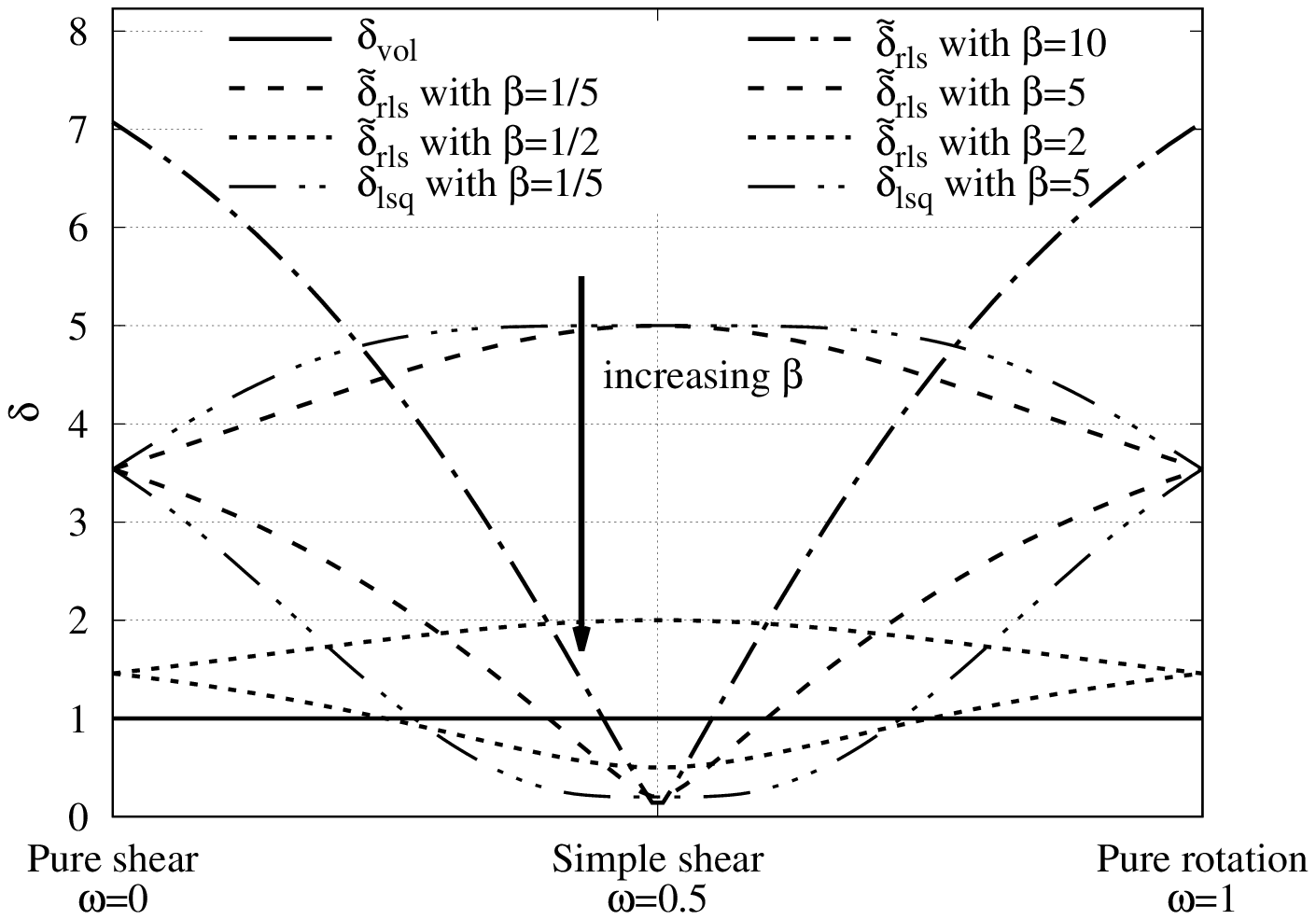}
}
\caption{Comparison between $\FLprls$ and $\FLvol$ for the simple 2D
  flow defined in Eq.(\ref{simpleflow}) with different values of
  $\beta=\{1/5,1/2,2,5,10\}$.}
\label{simple_flows}
\end{figure}

\mbigskip

Finally, results obtained for the following \Xavi{2D mesh and flow
  topology}
\begin{equation}
\label{simpleflow}
\DelTen = \Xavi{\left( \begin{array}{cc} \Dx & 0 \\ 0 & \Dy \end{array} \right)} = \left( \begin{array}{cc} \beta & 0 \\ 0 & \beta^{-1} \end{array} \right) , \hspace{7mm}
\G = \Xavi{\left( \begin{array}{cc} \partial_{x} \uvel & \partial_{y} \uvel \\ \partial_{\x} \vvel & \partial_{\y} \vvel \end{array} \right)} = \left( \begin{array}{cc} 0 & 1 \\ 1-2\omega & 0 \end{array} \right) ,
\end{equation}
\noindent are displayed in Figure~\ref{simple_flows}. \bef{Notice that
  the size of the control volume remains equal to unity; therefore,
  $\FLvol = 1$, regardless of the value of $\beta$.}  \aft{Since the
  control volume is fixed to unity, the Deardorff length scale,
  $\FLvol$, remains constant at $1$, independent of the value of
  $\beta$.} On the other hand, values of $\omega$ in
Figure~\ref{simple_flows} range from a pure shear flow ($\omega=0$) to
a simple shear flow ($\omega=1/2$), to a pure rotating flow
($\omega=1$). For the two limiting situations $\FLprls = \sqrt{(
  \beta^2 + \beta^{-2} )/2}$ whereas for $\omega=1/2$ it reads
$\FLprls = \beta^{-1}$. Recalling that, in this particular case,
$\Dx=\beta$ and $\Dy=\beta^{-1}$, $\FLprls = \sqrt{(\Dx^2 + \Dy^2)/2}$
for $\omega=0$ (pure shear) and $\omega=1$ (pure rotation), whereas
$\FLprls = \Dy$ for the simple shear flow with $\omega=1/2$. The
latter case closely corresponds to the typical quasi-2D, grid-aligned
flow observed in the initial region of a shear layer. As expected, the
computed $\FLprls$ (also the $\FLrls$ length scale) matches the grid
size in the direction perpendicular to the shear layer. The purely
rotational flow ($\omega = 1$) is a special case of
Eq.~(\ref{Drls_rot}), where $\vortx=\vorty=0$ and $\vortz=1$. \aft{It
  is worth noting that Figure~\ref{simple_flows} also includes results
  for the $\FLlsq$ length scale~\cite{TRI16-CharLength}, defined in
  Eq.~(\ref{DeltaLsq}). This length scale yields the same values as
  $\FLprls$ for pure shear ($\omega = 0$), simple shear flow ($\omega
  = 1/2$), and pure rotation ($\omega = 1$). Differences arise in the
  transitions between these regimes: specifically, $\FLlsq$ exhibits a
  wider plateau around $\omega = 1/2$, whereas $\FLrls$ transitions
  more rapidly. Similar behavior is observed for other values of
  $\beta$, with the differences between $\FLlsq$ and $\FLprls$
  remaining qualitatively the same.}

\begin{figure}[!t]
\centering{
  \includegraphics[height=0.6969\textwidth,angle=-90]{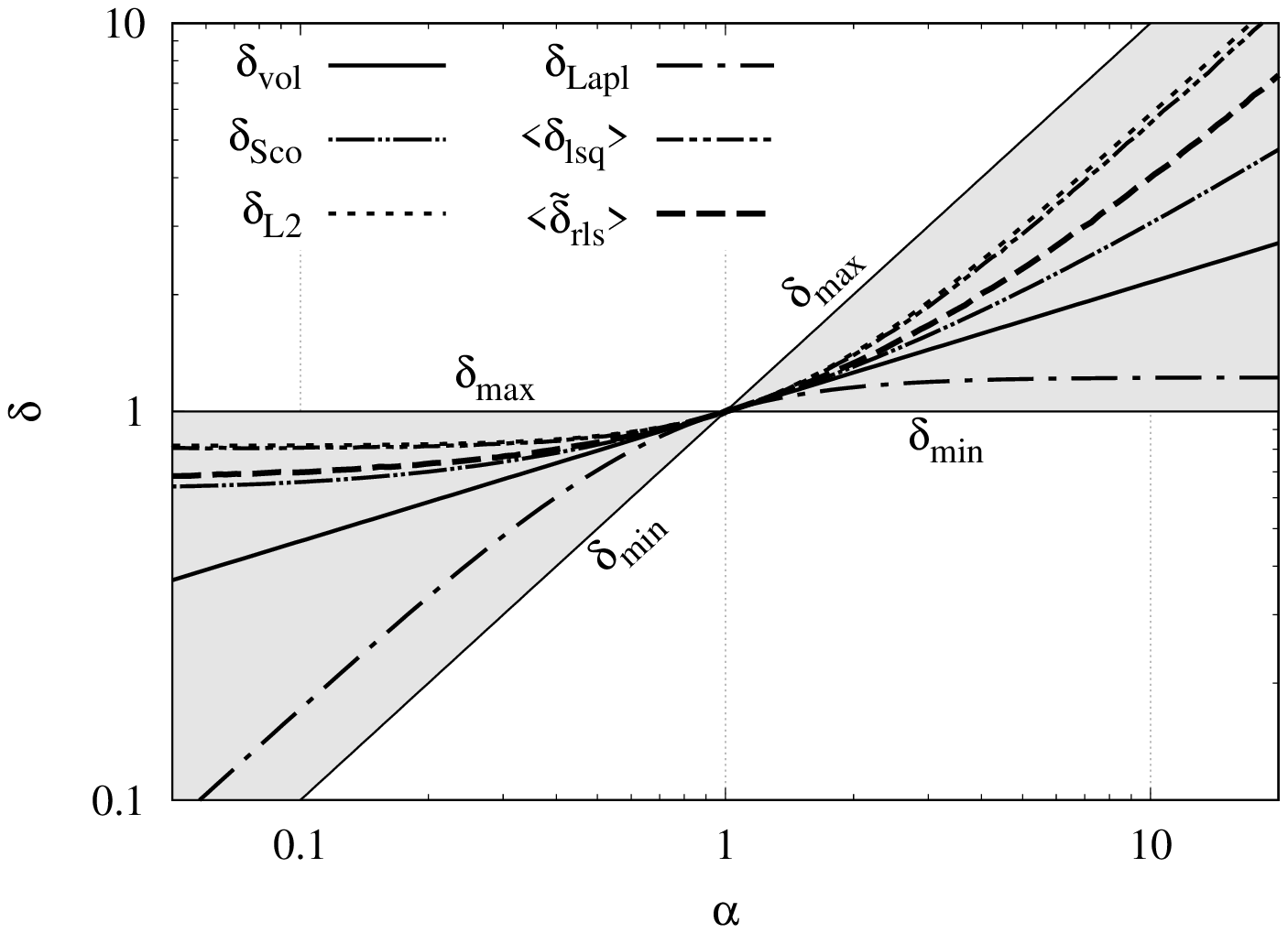}
}
\caption{Scaling of different definitions of $\Flength$ for a
  Cartesian mesh with \XaviJesus{$\Dx=\Dy=1$ and
    $\Dz=\alpha$}. Average results of $\FLlsq$ and $\FLprls$ have been
  obtained from a large enough sample of random traceless velocity
  gradient tensors.}
\label{scaling}
\end{figure}

\mbigskip

\bef{In order to study in more detail the effect of mesh anisotropies
  for different definitions of $\Flength$,} \aft{To gain deeper
  insight into the impact of mesh anisotropy on various $\Flength$
  definitions,} let us consider a Cartesian mesh with $\Dx=\Dy=1$ and
$\Dz=\alpha$. In this case, the geometry-dependent definitions of
$\Flength$ result in
\begin{align}
\begin{split}
\FLvol &= \alpha^{1/3}, \hspace{5mm} \FLSco = f(\min(\alpha,\alpha^{-1}),\min(1,\alpha^{-1})) \alpha^{1/3}, \\
\FLmax &= \max(1,\alpha), \hspace{5mm} \FLLtwo = \sqrt{\frac{2+\alpha^2}{3}}, \hspace{5mm} \FLLapl = \sqrt{\frac{3\alpha^2}{2\alpha^2+1}} .
\end{split}
\end{align}
\noindent where $f(a_1,a_2)$ is the correcting function given in
Eq.(\ref{DeltaScotti}). These functions are displayed in
Figure~\ref{scaling} using a log-log scale. Values of $\alpha > 1$
correspond to pencil-like meshes ($\Dx = \Dy \ll \Dz$) whereas values
of $\alpha < 1$ correspond to pancake-like meshes ($\Dx \ll \Dy =
\Dz$). The averaged values of $\FLlsq$ and $\FLprls$ are also
displayed; they have been obtained from a sufficiently large sample of
random traceless velocity gradient tensors, $\G$. Among all the
geometry-dependent definitions, the $\FLLtwo$ given in
Eq.(\ref{DeltaL2}) is by far the closest to $\avgtime{\FLlsq}$ for the
mesh considered here. In fact, for simple flow configurations such as
pure shear or pure rotation, $\FLlsq$ simplifies to $\FLLtwo$. On the
other hand, $\FLprls$ falls between $\FLLtwo$ and the correction
proposed by Scotti~\etal~\cite{SCO93} (see
Eq.\ref{DeltaScotti}). Finally, $\FLLapl$ is the only definition that
predicts values of $\Flength$ smaller than the classical Deardorff
length scale, $\FLvol$.

\begin{figure}[!t]
\centering{
  \includegraphics[angle=-90,width=0.63\textwidth]{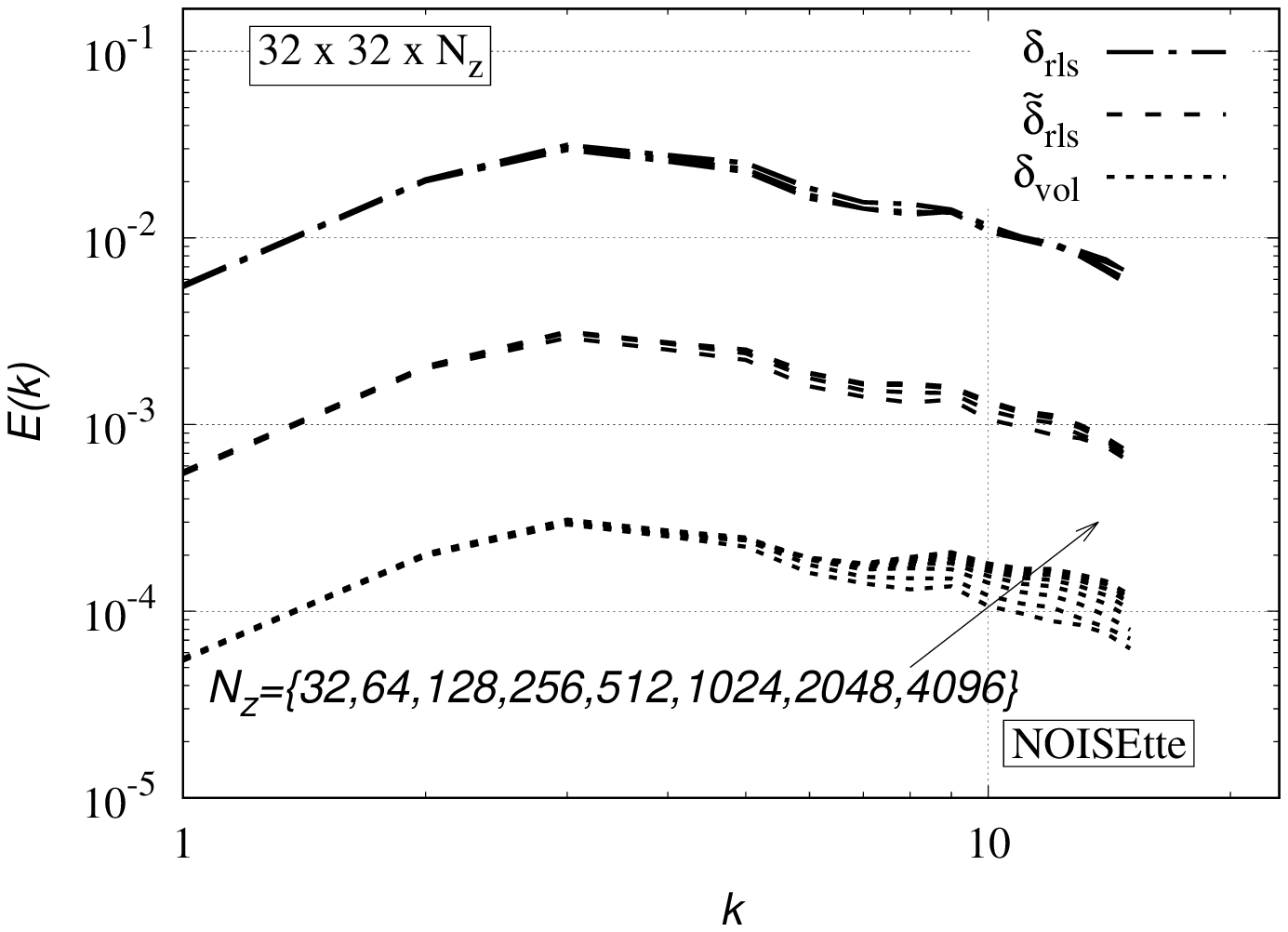}
  \includegraphics[angle=-90,width=0.63\textwidth]{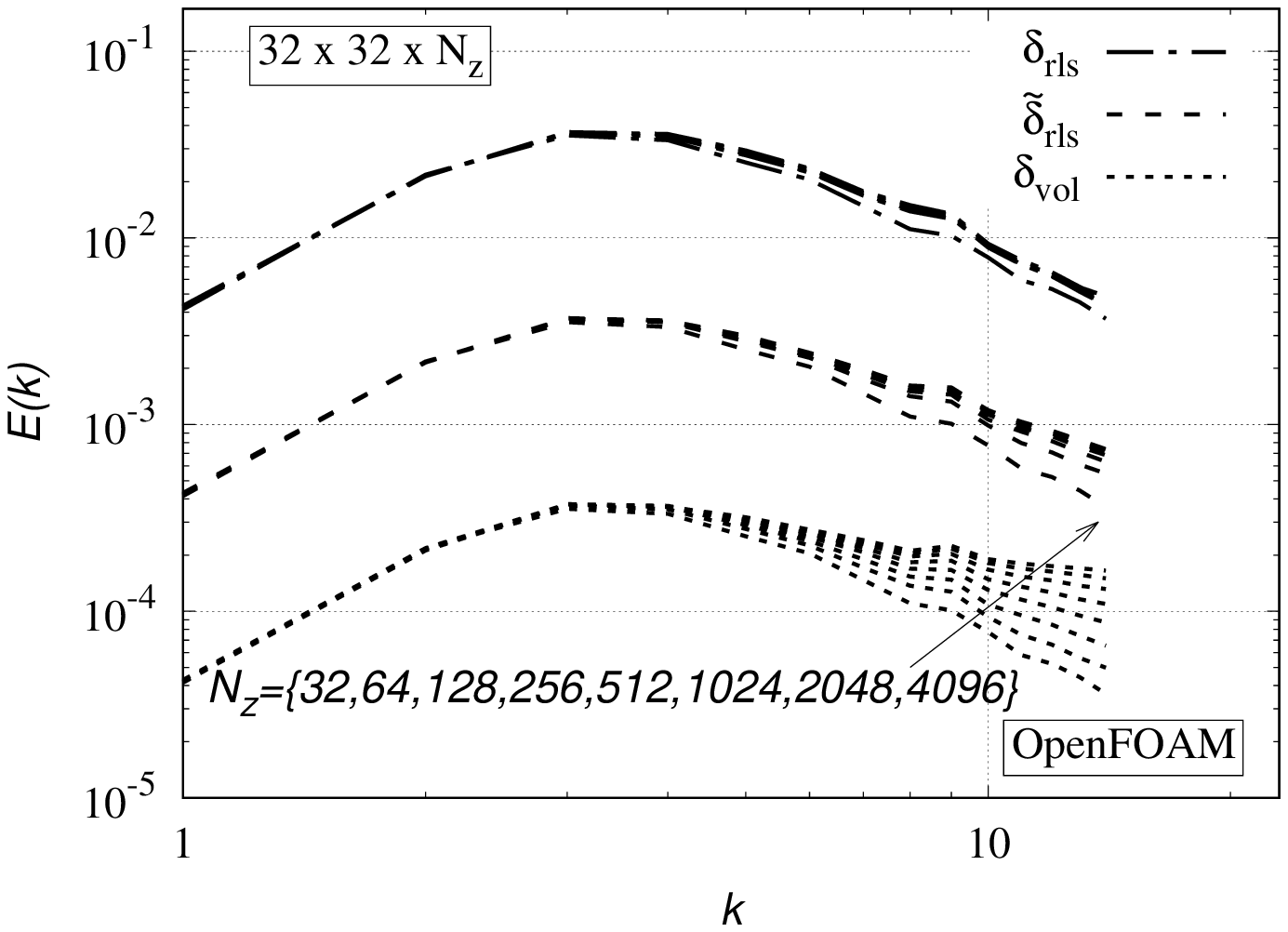}
}
\caption{Energy spectra for decaying isotropic turbulence
  corresponding to the Comte-Bellot and Corrsin experimental
  set-up~\cite{COM71}. LES results were obtained using the Smagorinsky
  model on a set of anisotropic meshes with pancake-like control
  volumes, employing two different codes: the in-house NOISEtte code
  (top) and OpenFOAM (bottom). Results obtained with the novel
  definitions of $\FLrls$ and $\FLprls$ respectively proposed
  in~Eqs.~(\ref{DeltaRLS}) and~(\ref{DeltaPRLS}) are compared with the
  classical definition proposed by Deardorff, given in
  Eq.~(\ref{DeltaDeardorff}). For clarity, the results obtained with
  $\FLprls$ and $\FLvol$ are shifted down one and two decades,
  respectively.}
\label{CBCresults_pancake}
\end{figure}

\begin{figure}[!t]
\centering{
  \includegraphics[angle=-90,width=0.63\textwidth]{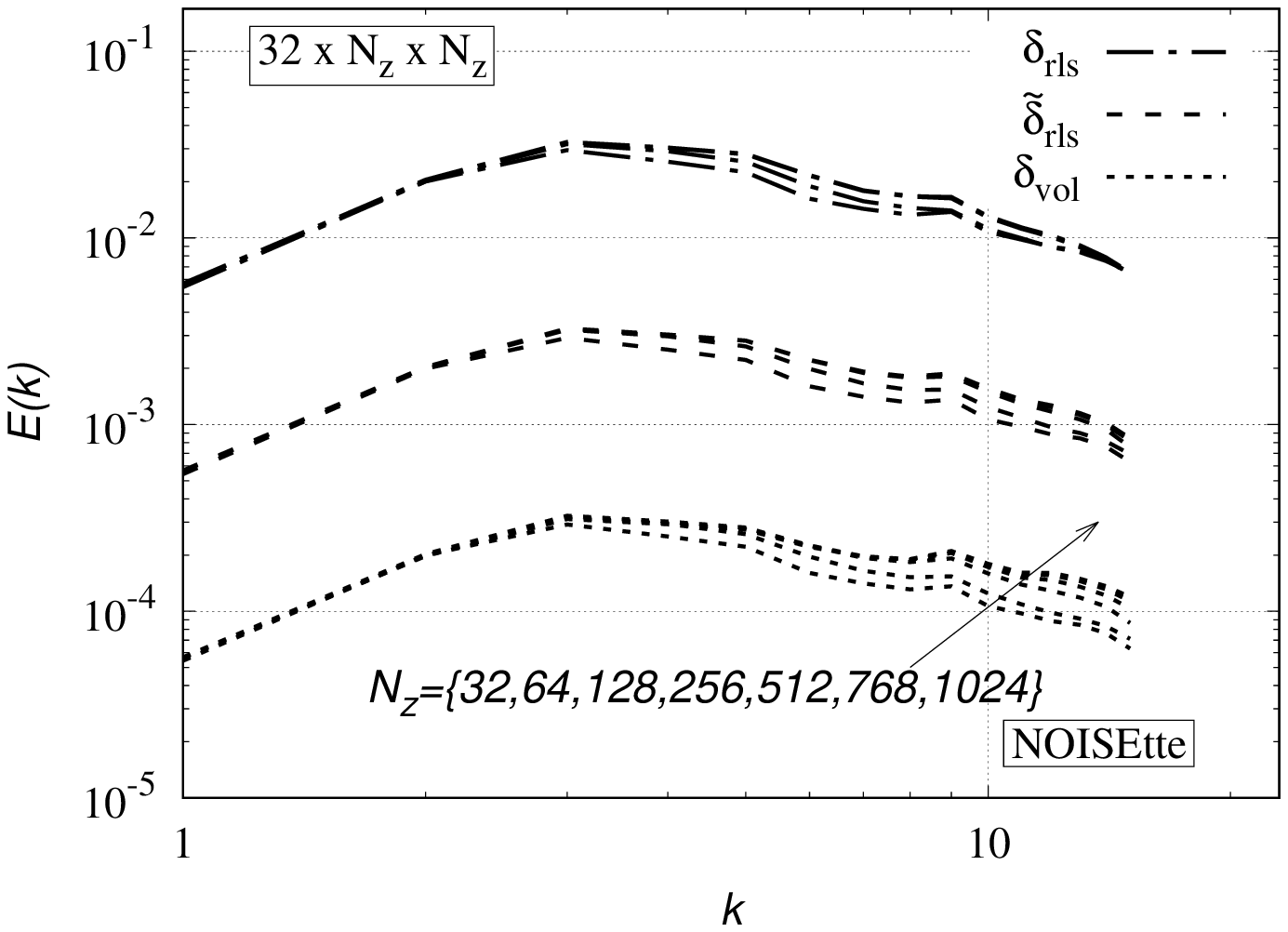}
  \includegraphics[angle=-90,width=0.63\textwidth]{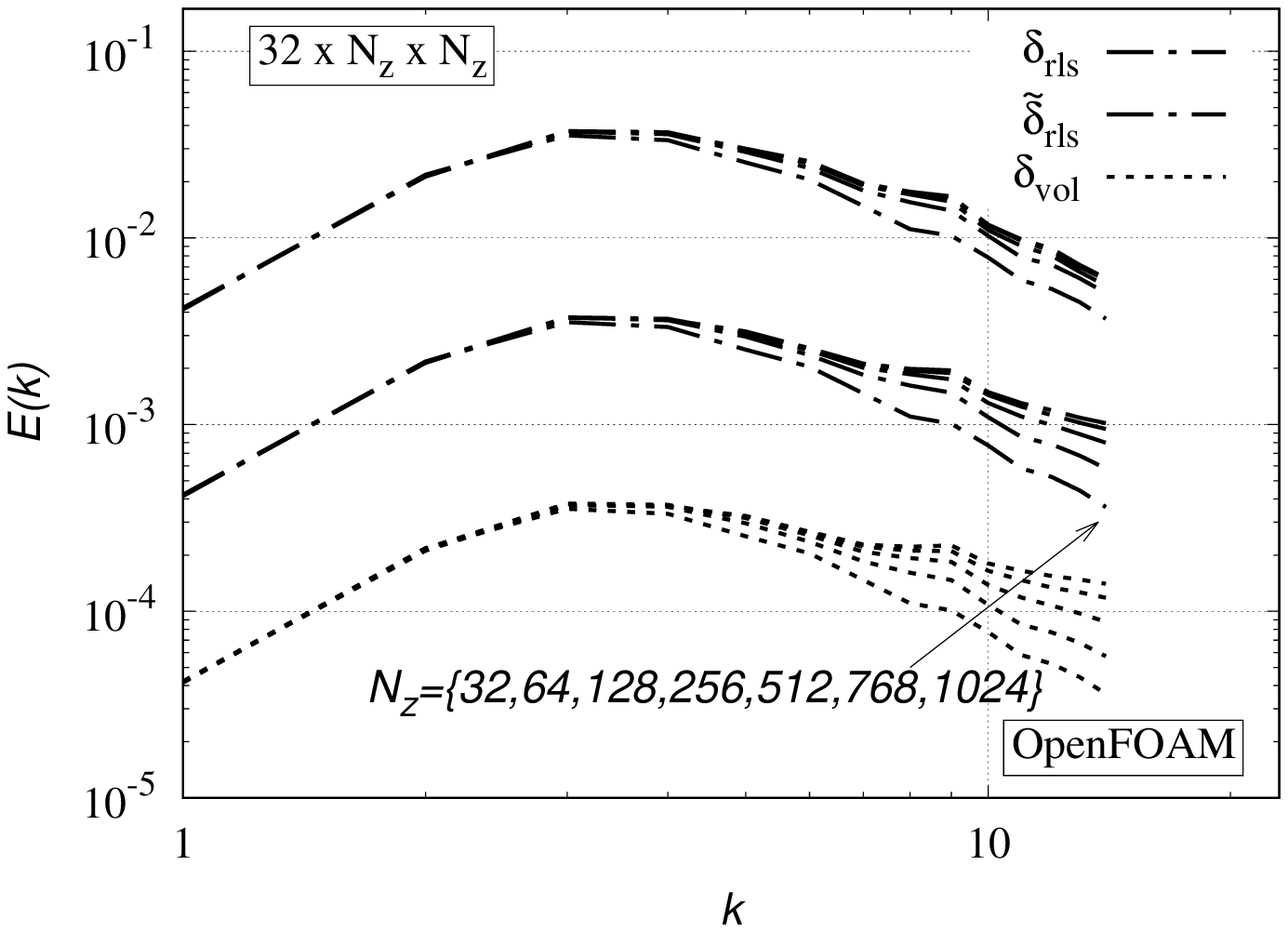}
}
\caption{Same as \Xavi{Figure~\ref{CBCresults_pancake}} but for
  pencil-like meshes.}
\label{CBCresults_pencil}
\end{figure}


\subsection{Isotropic turbulence on anisotropic grids}

Finally, the numerical simulation of decaying isotropic turbulence is
chosen to test the novel definitions of the subgrid characteristic
length scale, $\FLrls$ and $\FLprls$, proposed
in~Eqs.~(\ref{DeltaRLS}) and~(\ref{DeltaPRLS}), respectively. The
configuration corresponds to the classical experiment of Comte-Bellot
and Corrsin~\cite{COM71} (hereafter denoted as CBC) using the grid
turbulence with a size of $M=5.08cm$ and a free-stream velocity
$U_{0}=10 m/s$. The Taylor micro-scale Reynolds number at $t
U_{0}/M=42$ (initial state) is $Re_{\lambda} = u_{rms} \lambda / \nu =
71.6$ with $u_{rms}=22.2cm/s$ and decreases to $60.6$ at $tU_{0}/M =
171$ (third stage).

\mbigskip

The results are non-dimensionalized with the reference length
$L_{ref}=11M/(2\pi)$ and reference velocity $u_{ref}=\sqrt{3/2}
u_{rms} |_{t U_{0}/M=42}$. The energy spectrum of the initial field at
$t U_{0}/M=42$ matches the CBC experimental data. All subsequent
results are presented at $t U_{0}/M = 98$, which corresponds to the
second stage of the CBC experimental data.

\mbigskip

Simulations are carried out using two different codes: namely, the
in-house NOISEtte code~\cite{GOR22,ABA24} and OpenFOAM. NOISEtte is
based on high-accuracy Edge-Based Reconstruction (EBR) finite-volume
schemes~\cite{ABA16}. On translationally invariant (structured)
meshes, EBR schemes are equivalent to high-order finite-difference
methods, reaching up to sixth-order accuracy. \XaviAlex{For LES
  simulations, central-difference EBR4 was used to minimize artificial
  numerical dissipation.} On the other hand, OpenFOAM simulations were
performed using OpenFOAM~v2312, with appropriate modifications to the
{\it decayIsoTurb} tutorial. The solver used for all \XaviJesus{length
  scales} was {\it pimpleFoam}, employing two internal correctors and
a single outer corrector. The selected time integration scheme was the
implicit second-order {\it "Backward"} scheme, with a fixed time step
ensuring $CFL < 1$ for cases up to $32 \times 32 \times 1024$, while a
more stringent condition of $CFL < 0.5$ was applied for finer
meshes. The linear scheme was used for the gradient, divergence, and
Laplacian operators to minimize numerical diffusion. The setup for
these cases, along with all necessary files to reproduce the obtained
results in OpenFOAM (including modifications to the {\it
  TurbulenceModel} library) are publicly
available~\cite{RUA25-GitHub}. For both codes, the initial velocity
fields were obtained by interpolating a velocity field on a $64 \times
64 \times 64$ with an energy spectrum corresponding to the CBC initial
spectrum.

\mbigskip

LES results have been carried out for a set of (artificially)
stretched meshes using the Smagorinsky model with constant equal to
$C_S=0.17$ (NOISEtte) and $C_S=0.21$ (OpenFOAM). These constants have
been calibrated for the $32 \times 32 \times 32$ mesh and kept
constant for the rest of simulations. In particular, results for
pancake-like meshes with dimensions $32 \times 32 \times \Nz$, where
$\Nz = \{32, 64, 128, 256, 512, 1024, 2048, 4096\}$, are presented in
Figure~\ref{CBCresults_pancake}. As expected, when using Deardorff’s
classical definition (see Eq.\ref{DeltaDeardorff}), the results tend
to diverge as $\Nz$ increases. This occurs because $\FLvol$ approaches
zero with increasing $\Nz$, effectively disabling the SGS model. In
contrast, the newly proposed $\FLrls$ and $\FLprls$ avoid this issue,
and the results rapidly converge for larger values of $\Nz$. These
trends, observed for both codes, suggest that they effectively reduce
the influence of mesh anisotropies on the SGS model performance.

\mbigskip

\Xavi{A similar trend is observed in Figure~\ref{CBCresults_pencil}
  for pencil-like meshes composed of $32 \times \Nz \times \Nz$ grid
  cells, where $\Nz=\{32,64,128,256,512,768,1024\}$. In this case,
  $\FLvol$ scales as ${\mathcal O}(\Dz^{4/3})$ compared to the
  ${\mathcal O}(\Dz^{2/3})$ scaling for the pancake-like meshes,
  causing the eddy-viscosity model (see Eq.\ref{eddyvis_template}) to
  switch off even more quickly for increasing values of
  $\Nz$. Moreover, in this case, the numerical artifact affects a
  broader range of wavenumbers, whereas for pancake-like meshes, the
  impact is mostly confined to the smallest resolved scales (see
  Figure~\ref{CBCresults_pancake}). In contrast, LES results obtained
  with $\FLrls$ and $\FLprls$ show convergence as $\Nz$
  increases. However, more noticeable variations are observed for the
  first three mesh resolutions, \ie~$\Nz=\{32,64,128\}$, compared to
  the pancake-like cases. This slower convergence may be due to the
  fact that, in pencil-like meshes, finer resolution is achieved in
  two spatial directions instead of one, meaning that more physical
  scales are being resolved. As a result, the SGS model plays a
  reduced role, and the observed differences are likely related to the
  natural convergence due to grid refinement. In any case, the newly
  proposed subgrid characteristic length scales, $\FLrls$ and
  $\FLprls$, significantly mitigate the artificial effects caused by
  mesh anisotropies and ensure more consistent convergence behavior
  compared to the classical Deardorff approach.}

\begin{figure}[!t]
\centering{
  \includegraphics[angle=-90,width=0.63\textwidth]{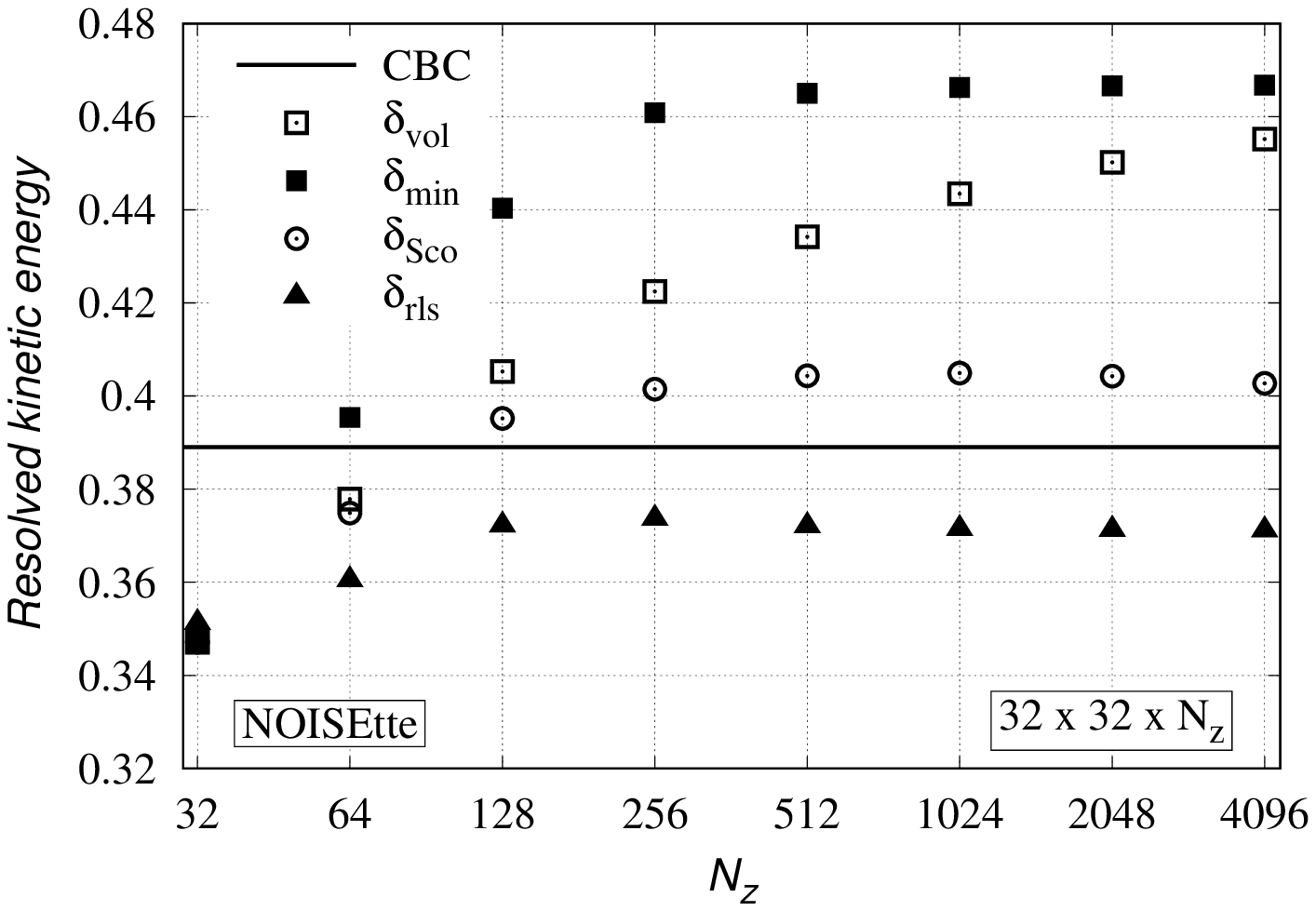}
  \includegraphics[angle=-90,width=0.63\textwidth]{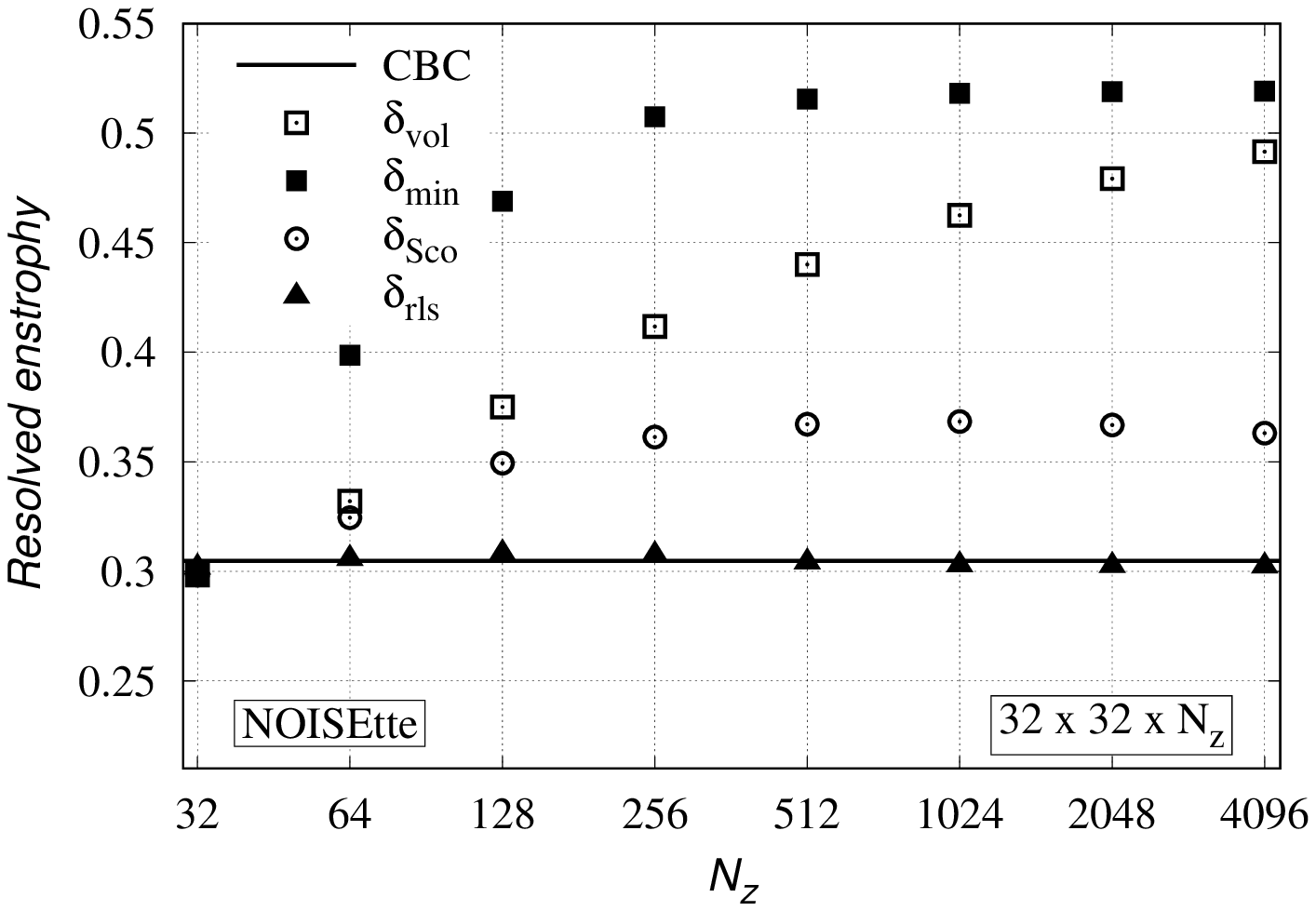}
}
\caption{Resolved kinetic energy (top) and enstrophy (bottom) as a
  function of the number of grid points, $\Nz$, for decaying isotropic
  turbulence corresponding to the experiment of Comte-Bellot and
  Corrsin~\cite{COM71}. LES results have been obtained using the
  Smagorinsky model for a set of anisotropic meshes with pancake-like
  control volumes, \ie~$32 \times 32 \times \Nz$. Results correspond
  to definitions of $\Flength$ that depend solely on the mesh
  geometry: $\FLvol$, $\FLmin$, $\FLSco$ and the novel definition
  $\FLrls$.}
\label{CBCconvergence_MeshDependent}
\end{figure}

\begin{figure}[!t]
\centering{
  \includegraphics[angle=-90,width=0.63\textwidth]{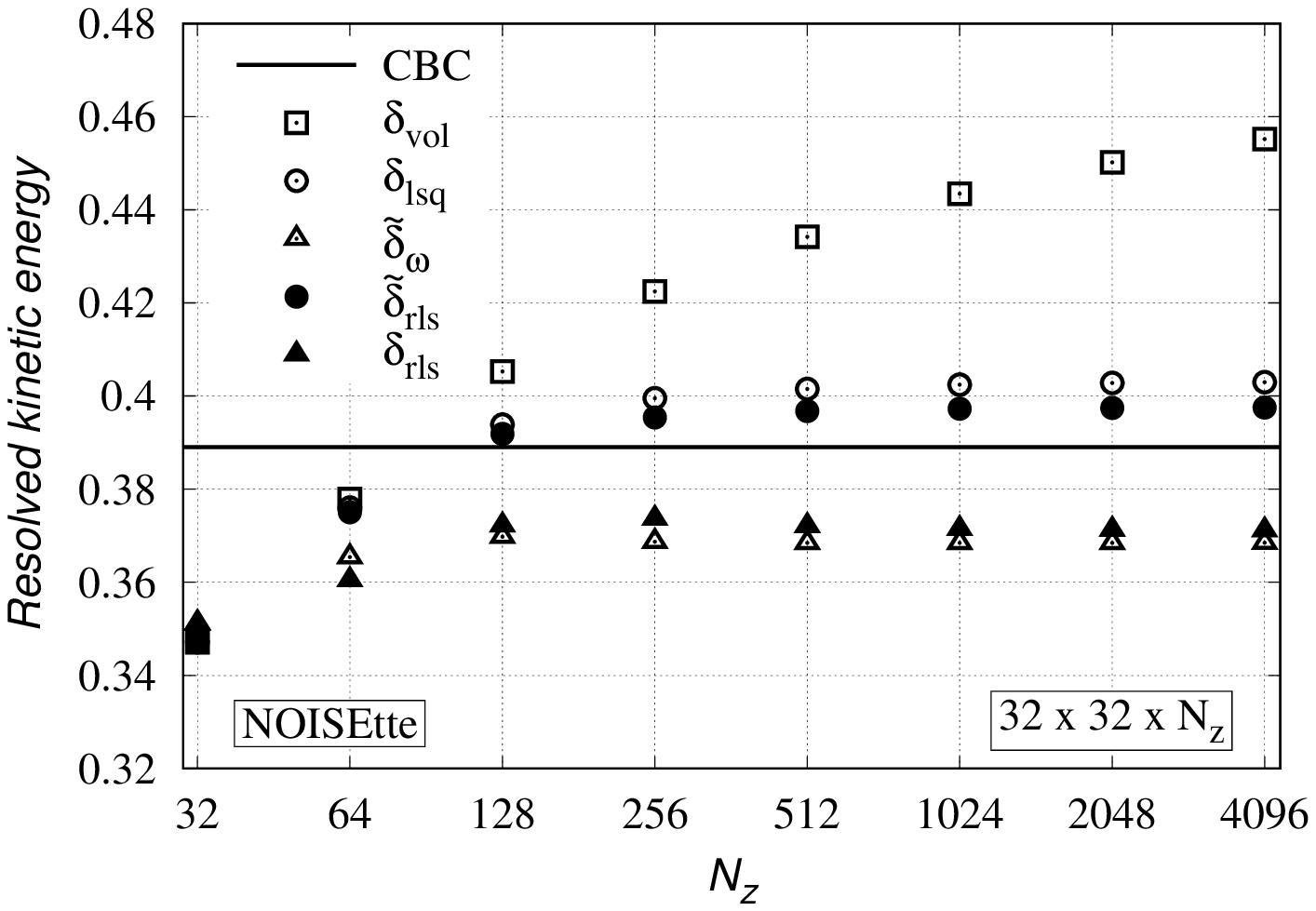}
  \includegraphics[angle=-90,width=0.63\textwidth]{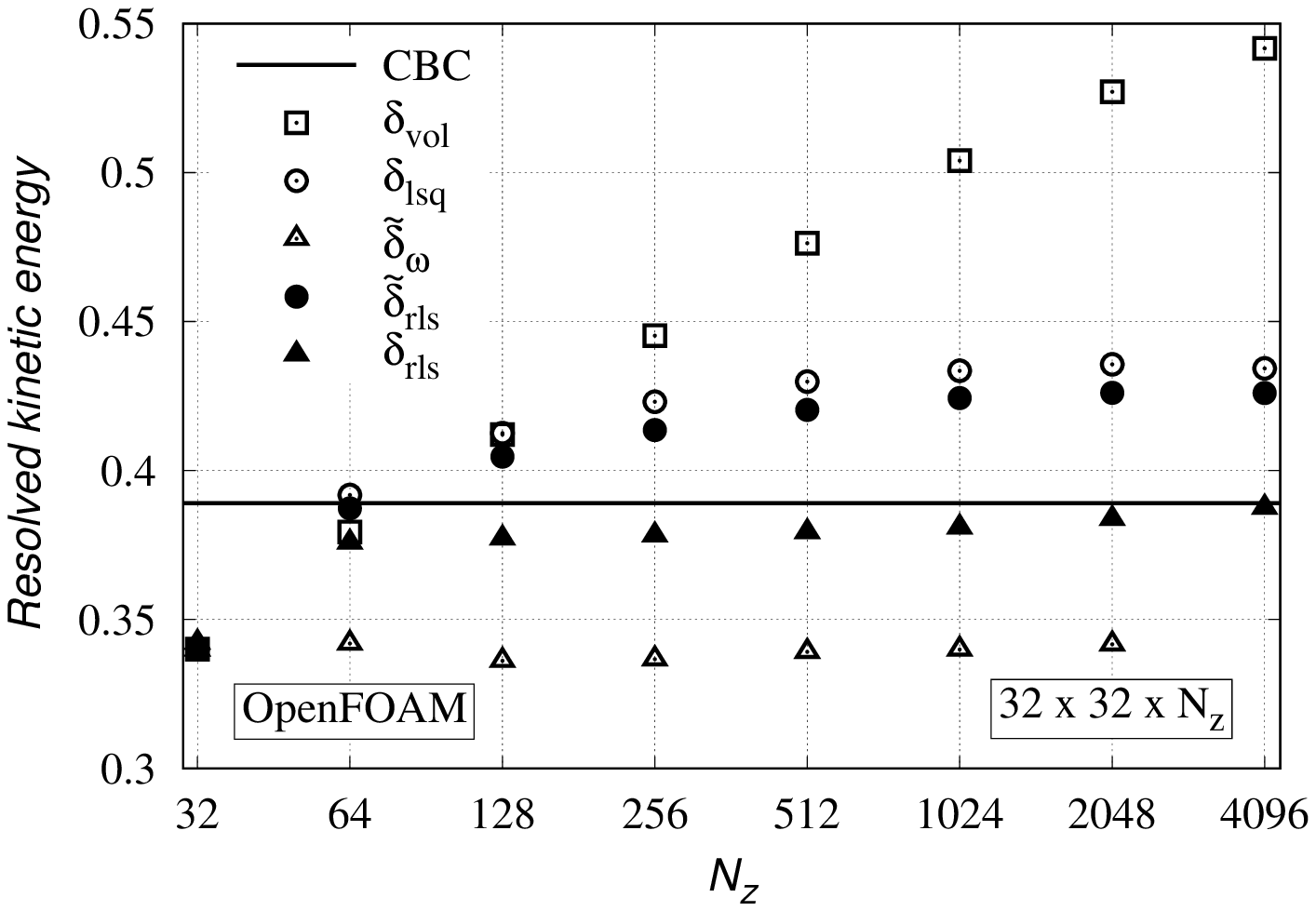}
}
\caption{Same as Figure~\ref{CBCconvergence_MeshDependent} (top). In
  this case, results were obtained with two different codes: the
  in-house NOISEtte code (top) and OpenFOAM (bottom). Moreover, in
  addition to $\FLvol$ and $\FLrls$ (also shown in
  Figure~\ref{CBCconvergence_MeshDependent}), results include
  definitions of $\Flength$ that depend on the flow topology:
  $\FLlsq$, $\FLMoc$, and the newly proposed definition, $\FLprls$.}
\label{CBCconvergence_KE_FlowDependent}
\end{figure}

\begin{figure}[!t]
\centering{
  \includegraphics[angle=-90,width=0.63\textwidth]{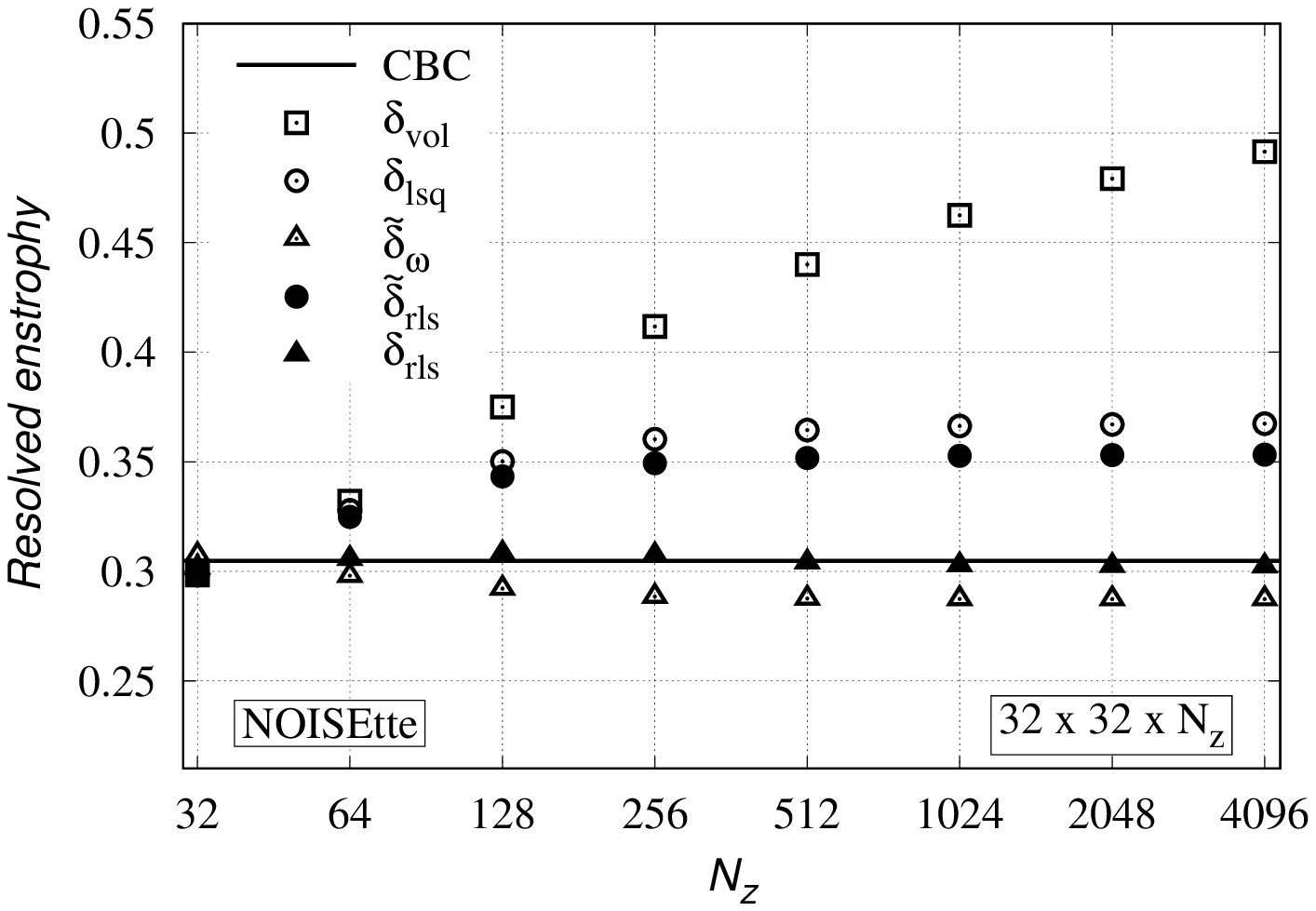}
  \includegraphics[angle=-90,width=0.63\textwidth]{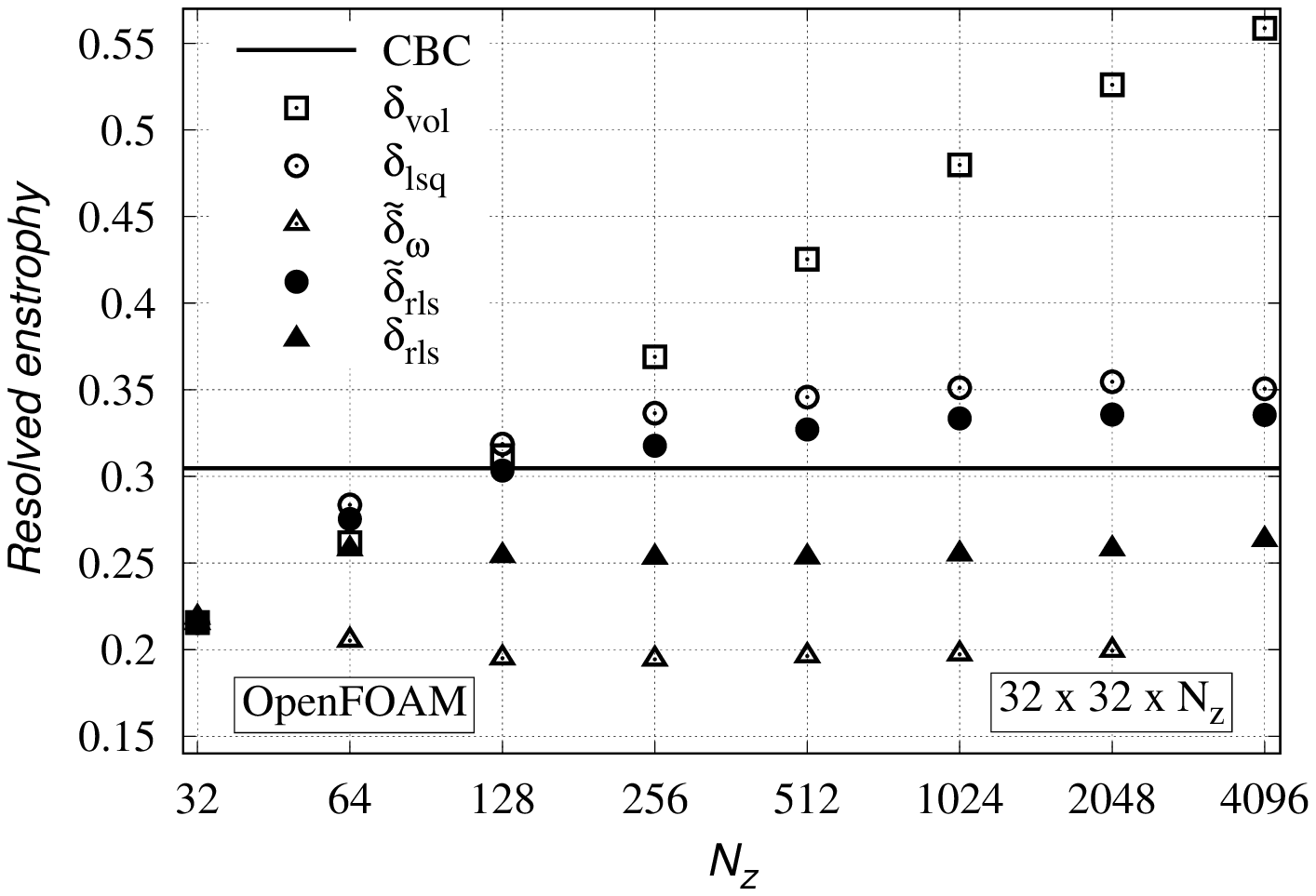}
}
\caption{Same as Figure~\ref{CBCconvergence_MeshDependent}
  (bottom). In this case, results were obtained with two different
  codes: the in-house NOISEtte code (top) and OpenFOAM
  (bottom). Moreover, in addition to $\FLvol$ and $\FLrls$ (also shown
  in Figure~\ref{CBCconvergence_MeshDependent}), results include
  definitions of $\Flength$ that depend on the flow topology:
  $\FLlsq$, $\FLMoc$, and the newly proposed definition, $\FLprls$.}
\label{CBCconvergence_ENSTROPHY_FlowDependent}
\end{figure}

To gain deeper insight into the impact of mesh anisotropies on
different definitions of $\Flength$, two key physical quantities have
been analyzed: the resolved kinetic energy and the resolved
enstrophy. \aft{References values are computed by integrating the
  second stage of the CBC data~\cite{COM71} up to the cut-off length
  scale defined by our LES resolution. This leads to a reference value
  for the resolved turbulent kinetic energy of approximately $0.39$,
  which corresponds to around $67\%$ of the total turbulent kinetic
  energy.} Results displayed in
Figures~\ref{CBCconvergence_MeshDependent},~\ref{CBCconvergence_KE_FlowDependent}
and~\ref{CBCconvergence_ENSTROPHY_FlowDependent} have been obtained
with the same pancake-like meshes, \ie~$32 \times 32 \times \Nz$ grid
cells where $\Nz=\{32,64,128,256,512,1024,2048,4096\}$. In this case,
we have firstly compared definitions that depend solely on the mesh
geometry and not on the flow topology (see
Figure~\ref{CBCconvergence_MeshDependent}). This analysis was
performed exclusively for NOISEtte. Namely, in addition to the new
approach, $\FLrls$, and the Deardorff length scale, $\FLvol$, two
other definitions have been tested: the definition proposed by
Scotti~\etal~\cite{SCO93}, $\FLSco$, given in Eq.~(\ref{DeltaScotti}),
and $\FLmin = \min ( \Dx , \Dy , \Dz ) $, which establishes a lower
bound for the other definitions. It is important to note that $\FLSco$
is chosen for this comparison because, among all the definitions
reviewed in Section~\ref{intr} that rely solely on the geometrical
properties of the mesh (see property {\bf P3} in
Table~\ref{properties_Delta}), it is the length scale that yields the
best results. As expected, the results obtained with $\FLvol$ get
closer and closer to those obtained with $\FLmin$ as $\Nz$ increases,
since in both cases, the SGS model tends to switch off. Remarkably,
the new definition $\FLrls$ clearly outperforms other ones, including
Scotti's length scale, $\FLSco$. As explained in Section~\ref{intr},
this definition of $\Flength$ was proposed as a correction of the
Deardorff definition, $\FLvol$, for anisotropic meshes with the
assumption of an isotropic turbulent regime. Therefore, it is not
surprising that $\FLSco$ behaves very robustly for a simulation of
decaying homogeneous isotropic turbulence. Furthermore, it is very
remarkable the robustness of the new definition in the enstrophy
results displayed in Figure~\ref{CBCconvergence_MeshDependent}
(bottom). Notice that enstrophy, which is the L2-norm, $\enstrophy
\equiv || \vort ||^2 = \int_\Omega | \vort |^2 \ud \Omega$ , of the
vorticity vector, $\vort \equiv \nabla \times \vel$, reads
$\enstrophy_k = k^2 E_k$ in Fourier space. Therefore, enstrophy
magnifies errors in the tail of the kinetic energy spectrum which is a
key parameter in the assessment of SGS models.

\mbigskip

The analysis is further extended in
Figures~\ref{CBCconvergence_KE_FlowDependent}
and~\ref{CBCconvergence_ENSTROPHY_FlowDependent} by considering length
scales that depend on the local flow topology; namely, $\FLlsq$,
$\FLMoc$ and the newly proposed $\FLprls$ and $\FLrls$. We selected
$\FLMoc$ because it was originally introduced as an improvement over
the definition by Chauvet~\etal~\cite{CHA07} given in
Eq.(\ref{Deltavort}). On the other hand, the definition proposed by
Shur\etal~\cite{SHU15}, $\FLSLA$, given in Eq.~(\ref{DeltaSLA}), has
not been considered in this analysis, as it is a modified version of
$\FLMoc$ specifically tailored to trigger the \KH instability in the
early stages of shear layers. Apart from this, notice that, although
the (face) evaluation of $\FLrls$ does not depend on the local flow
topology, its practical effect of the SGS models does
depend. Therefore, it is also considered in this analysis. Moreover,
for comparison, $\FLvol$ is also included, as it represents the
standard length scale used in the community. Again, it is remarkable
the performance of $\FLrls$ on highly anisotropic grids, showing even
more robust results than the best flow-dependent definitions such as
$\FLMoc$. This is specially clear in the enstrophy analysis displayed
in Figure~\ref{CBCconvergence_ENSTROPHY_FlowDependent}. Apart from
this, we can also observe that $\FLprls$ has a very similar behavior
as $\FLlsq$, constituting an excellent alternative to $\FLrls$ if the
SGS model is implemented following the standard approach. Finally, we
can conclude that the newly proposed approach $\FLrls$ displayed the
best results among all the existing length scale definitions. This is
even more remarkable when considering the simplicity of this new
approach compared with the complexity of other alternatives such as
$\FLMoc$ and $\FLSLA$.


\begin{figure}
\centering{
  \includegraphics[angle=-90,width=0.63\textwidth]{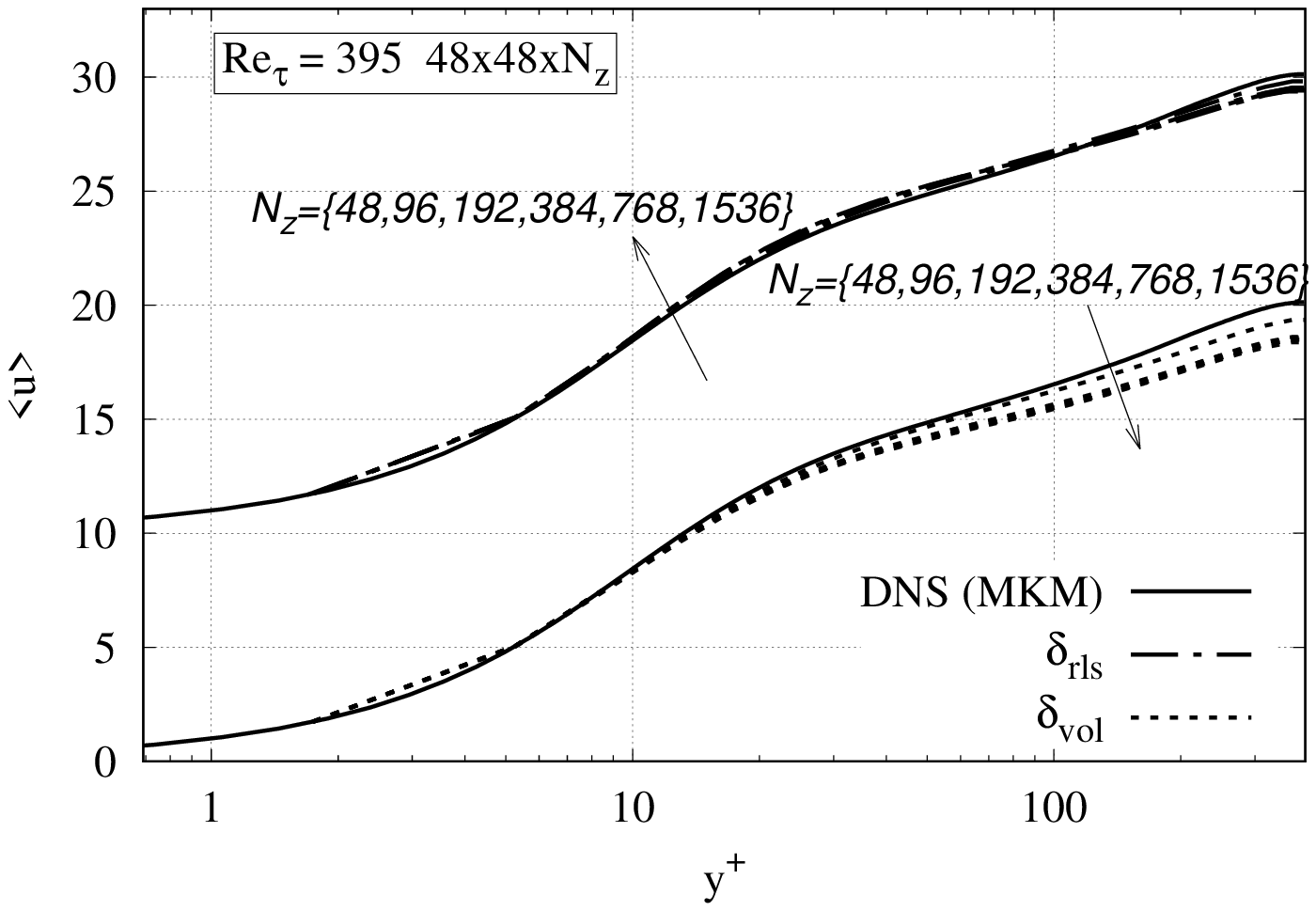}
}
\caption{\aft{Mean velocity, $\avgtime{\uvel}$, for a turbulent
    channel flow at $Re_{\tau}=395$ obtained for a set of anisotropic
    meshes $48 \times 48 \times \Nz$ with
    $\Nz=\{48,96,192,384,768,1536\}$. Solid lines corresponds to the
    DNS results by Moser~\etal~\cite{MOS99}. LES results obtained
    using the novel definition, $\FLrls$, are compared against those
    using $\FLvol$, with both employing the S3QR
    model~\cite{TRI14-Rbased}.  For clarity, the former results are
    shifted up.}}
\label{results_CF_avg}
\end{figure}

\begin{figure}
\centering{
\includegraphics[angle=-90,width=0.63\textwidth]{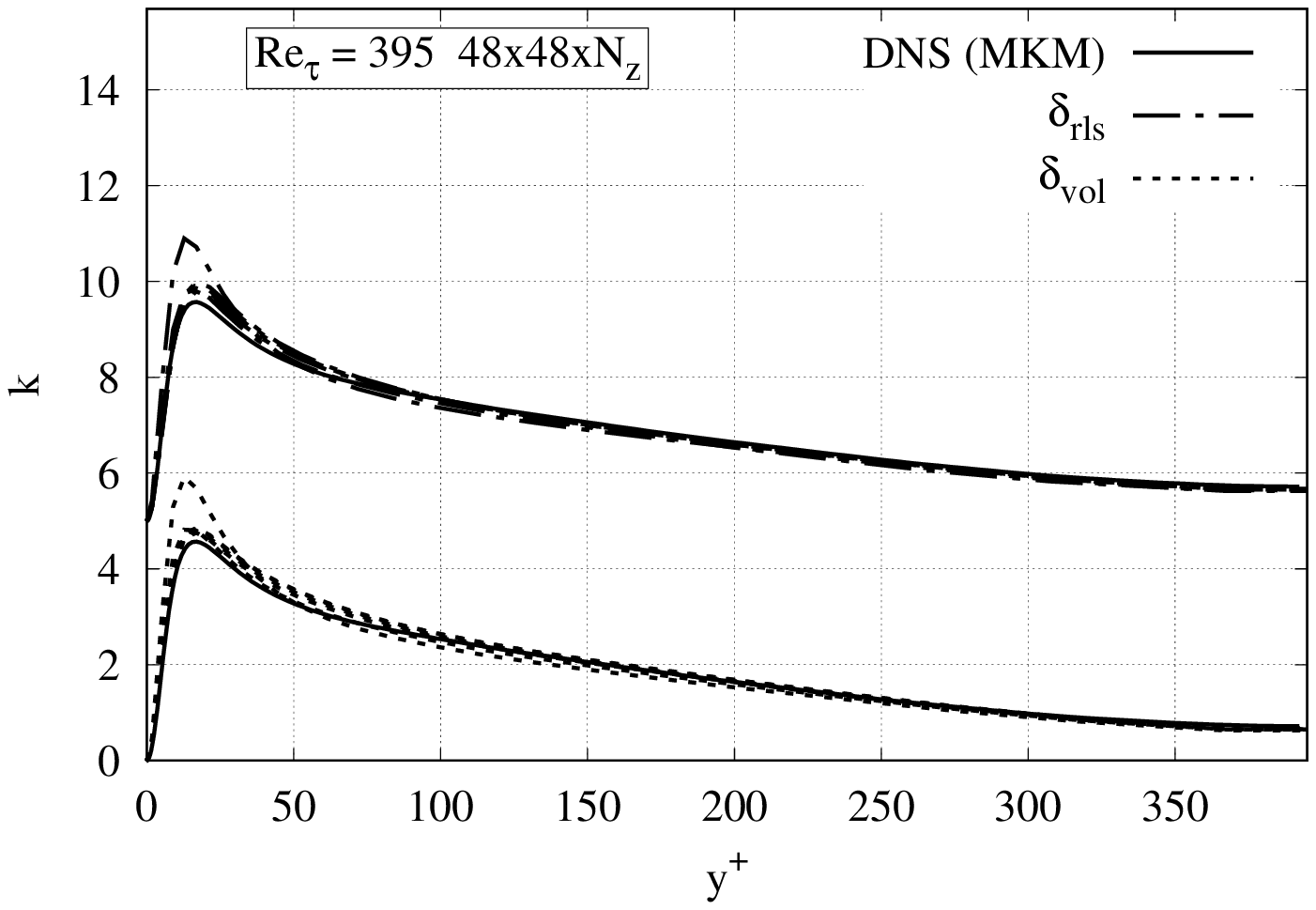}
\includegraphics[angle=-90,width=0.63\textwidth]{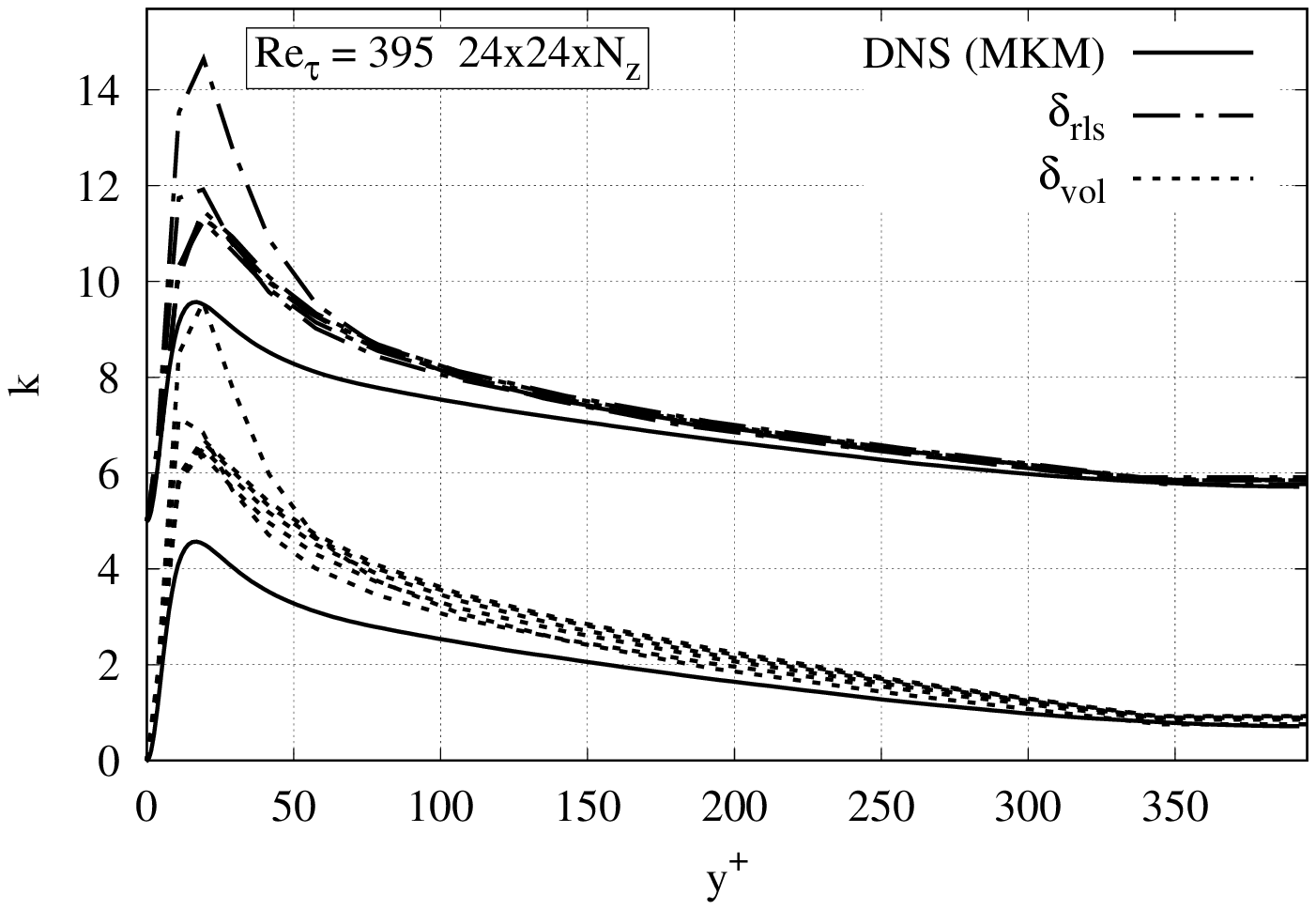}
}
\caption{\aft{Same as in Figure~\ref{results_CF_avg} but for the
    turbulent kinetic energy, $k$. In this case, two sets of
    anisotropic meshes have been analyzed. Top: $48 \times 48 \times
    \Nz$ with $\Nz=\{48,96,192,384,768,1536\}$. Bottom: $24 \times 24
    \times \Nz$ with $\Nz=\{24,48,96,192,384,768\}$.}}
\label{results_CF_k}
\end{figure}

\begin{figure}
\centering{
\includegraphics[angle=-90,width=0.63\textwidth]{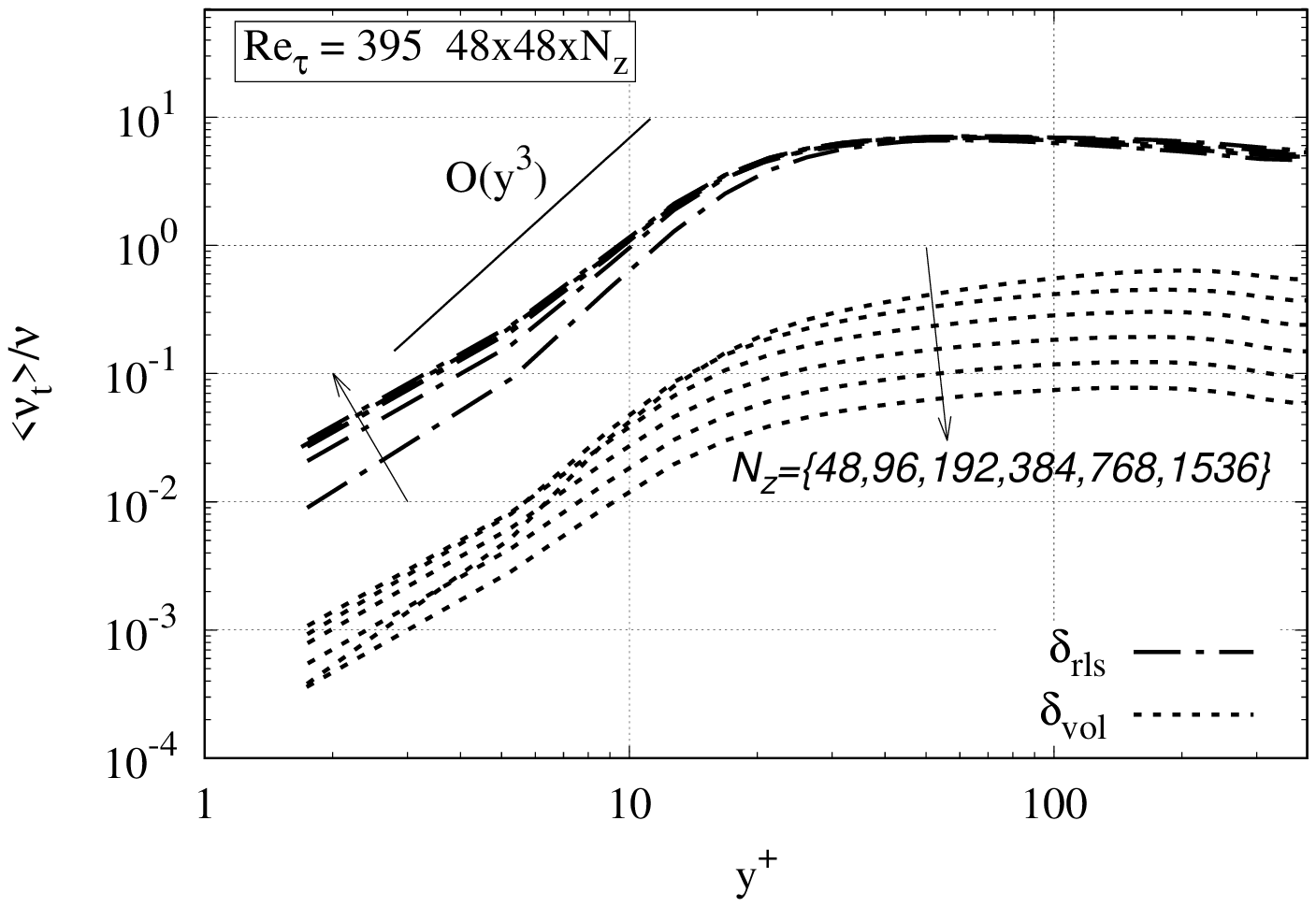}
\includegraphics[angle=-90,width=0.63\textwidth]{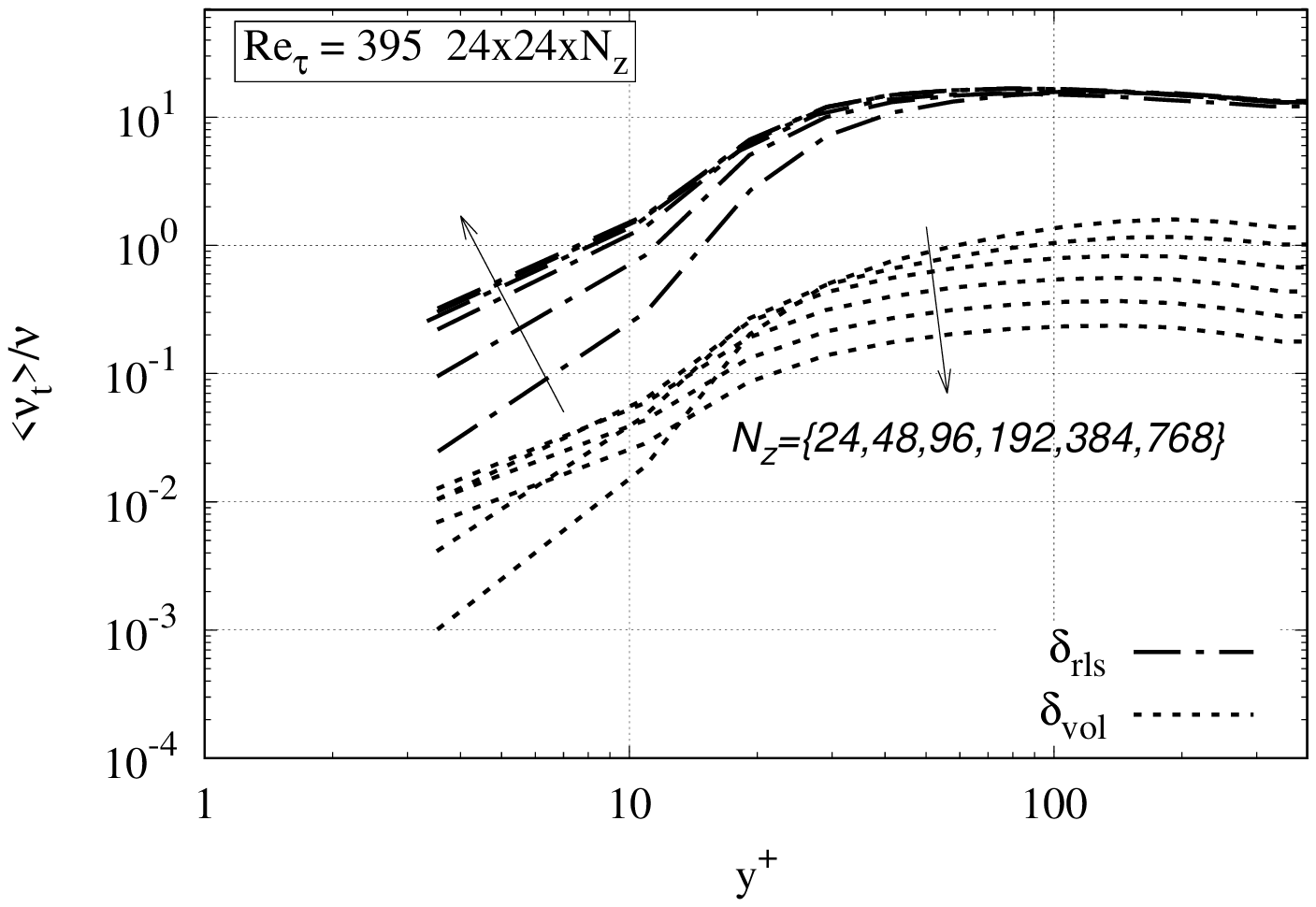}
}
\caption{\aft{Same as in Figure~\ref{results_CF_k} but for the ratio
    between the average turbulent viscosity, $\avgtime{\nut}$, and the
    molecular viscosity, $\nu$.}}
\label{results_CF_nut}
\end{figure}

\subsection{\aft{Turbulent channel flow}}

\aft{Finally, to assess the performance of the new definition $\FLrls$
  in the presence of walls, we consider a turbulent channel flow at
  $Re_\tau=395$. For clarity, this test compares only $\FLrls$, which
  has demonstrated superior results over other alternatives including
  $\FLprls$, against the standard length scale $\FLvol$. The
  simulations are performed using a code based on a
  symmetry-preserving finite-volume discretization~\cite{VER03} of the
  incompressible NS equations. The Smagorinsky model is not suitable
  for this case due to its inability to correctly capture the
  near-wall scaling of the turbulent viscosity~\cite{CHA86}, \ie~$\nut
  \propto y^3$. Therefore, it has been replaced by the S3QR model,
  proposed in a previous work~\cite{TRI14-Rbased}. Namely,
\begin{equation}
\label{S3QR} 
\nut^{S3QR} =  ( C_{s3qr} \Flength )^2 \Q_{\G\G\traspose}^{-1} \R_{\G\G\traspose}^{5/6} ,
\end{equation}
\noindent where $C_{s3qr} = 0.762$, $\Q_{\G\G\traspose}$ and
$\R_{\G\G\traspose}$ are the second and third invariants of the
symmetric second-order tensor $\G\G\traspose$.}

\mbigskip

\aft{Figure~\ref{results_CF_avg} shows the average velocity profiles
  obtained for a set of (artificially) refined meshes in the span-wise
  direction, with resolutions of $48 \times 48 \times \Nz$ and $\Nz =
  \{48, 96, 192, 384, 768, 1536\}$. Refinement is applied in the
  span-wise direction because the results are expected to be less
  sensitive to resolution in this direction compared to the
  stream-wise and wall-normal directions. LES results are compared
  with the DNS data by Moser~\etal~\cite{MOS99}, and the channel
  dimensions are set to match those of the DNS, \ie~$2\pi \times 2
  \times \pi$. The grid is uniform in both the stream-wise and
  span-wise directions, while the wall-normal grid points are
  distributed using a hyperbolic sine function given by
\begin{equation}
y_j = \sinh ( \gamma j / \Ny ) / \sinh ( \gamma / 2 )  \hspace{5mm} j=0,1,\dots, \Ny/2 .
\end{equation}
\noindent Here, $\Ny$ is the number of grid points in the wall-normal
direction, and the stretching parameter $\gamma$ is set to 7. The grid
points in the upper half of the channel are then obtained by
symmetry. With this distribution and $\Ny = 48$, the first off-wall
grid point is located at $y^+ \approx 1.75$, \ie~within the viscous
sublayer ($y^+ < 5$). The corresponding resolutions are $\Dx^+ \approx
51.7$ and $\Dz^+ \approx 25.9$ (for $\Nz=48$). Therefore, the mesh is
highly anisotropic near the wall, with, for instance, $\Dx^+ / \Dy^+
\approx 14.8$ in the first off-wall control volume.}

\mbigskip

\aft{As already observed in the homogeneous isotropic turbulence
  test-case, the results in Figure~\ref{results_CF_avg} confirm that
  the new definition of $\FLrls$ exhibits significantly greater
  robustness to mesh anisotropy. Notably, the mean velocity profile
  remains virtually unchanged across all mesh refinements when using
  the proposed length scale, whereas substantial variations are
  observed with the standard definition $\FLvol$. A similar trend is
  observed in Figure~\ref{results_CF_k} for the resolved turbulent
  kinetic energy, $k$, particularly in the bulk region, where the
  results obtained with $\FLrls$ show minimal dependence on the value
  of $\Nz$. This figure also includes results for a grid twice coarser
  on each spatial direction, \ie~$24 \times 24 \times \Nz$, with
  $\Nz=\{24, 48, 96, 192, 384, 768\}$, shown at the bottom of the
  Figure~\ref{results_CF_k}. Although the resolution is insufficient
  for accurate prediction, the robustness of the new length scale
  under such coarse conditions is particularly remarkable.}

\mbigskip

\aft{Finally, Figure~\ref{results_CF_nut} presents the ratio between
  the time-averaged turbulent viscosity, $\avgtime{\nut}$, and the
  molecular viscosity, $\nu$. These results provide valuable insight
  into the behavior of both length scale definitions. Specifically,
  $\FLrls$ consistently yields similar values of $\avgtime{\nut}$
  regardless of the resolution in the span-wise direction, while
  $\FLvol$ clearly tends to switch off the SGS model as $\Nz$
  increases. This contrasting behavior is especially evident in the
  bulk region of the channel. On the other hand, in the near-wall
  region, the $\nut$ tends to vanish towards the wall following the
  correct cubic behavior, \ie~$\nut \propto y^3$ (see
  Figure~\ref{results_CF_nut}, top). Notice that the number of grid
  points in the wall-normal direction is insufficient to properly
  capture this behavior for the coarsest mesh (see
  Figure~\ref{results_CF_nut}, bottom). In this region, noticeable
  differences persist for both $\FLrls$ and $\FLvol$ as the mesh is
  refined in the span-wise direction. At first sight, $\FLvol$ tends
  to predict lower values of $\avgtime{\nut}$ with increasing $\Nz$,
  while $\FLrls$ shows the opposite trend. Nevertheless, the values of
  $\avgtime{\nut}$ predicted by $\FLrls$ converge rapidly and
  monotonically with grid refinement.}

\mbigskip

\aft{In summary, the turbulent channel flow test confirms the
  robustness of the proposed length scale, $\FLrls$, in wall-bounded
  configurations with strong mesh anisotropy. The results show that
  $\FLrls$ provides consistent predictions for mean velocity,
  turbulent kinetic energy, and turbulent viscosity across a wide
  range of span-wise resolutions. In contrast, the standard definition
  $\FLvol$ exhibits significant sensitivity to mesh refinement,
  particularly in the bulk region where it tends to prematurely
  deactivate the SGS model. These findings further support the
  suitability of $\FLrls$ as a simple and reliable subgrid
  characteristic length for LES simulations, including near-wall
  regions.}


\section{Concluding remarks}

\label{conclusions}

\bef{In this paper, a novel approach for computing the subgrid
  characteristic length has been proposed to address the following
  research question:} \aft{This work proposes a novel definition of
  the subgrid characteristic length, $\Flength$, aimed at addressing
  the following research question:} {\it can we establish a simple,
  robust, and easily implementable definition of $\Flength$ for any
  type of grid that minimizes the impact of mesh anisotropies on the
  performance of SGS models for LES?} \bef{This is primarily motivated
  by the observation that, despite its known inaccuracies on highly
  anisotropic grids, the Deardorff definition, \ie~the cube root of
  the cell volume (see Eq.~\ref{DeltaDeardorff}), remains the most
  commonly used method in both academia and industry.} \aft{The
  motivation stems from the widespread use of the Deardorff definition
  (see Eq.~\ref{DeltaDeardorff}), \ie~the cube root of the cell
  volume, even though it is known to perform poorly on highly
  anisotropic grids.}

\mbigskip

In this context, we argue that the proposed {\it rational length
  scale}, $\FLrls$, defined in Eq.~(\ref{DeltaRLS}), is an excellent
alternative. It naturally emerges from the entanglement between
numerical discretization and LES filtering. Furthermore, its favorable
mathematical properties and simplicity indicate that it is a robust
and reliable choice, effectively reducing the impact of mesh
anisotropies on simulation accuracy. Specifically, it is locally
defined, frame-invariant, and well-bounded (see properties {\bf P1}
and {\bf P2} in Section~\ref{properties}), well-conditioned, and has a
very low computational cost (property {\bf P6}). Additionally, it is
suitable for unstructured meshes (property {\bf P4}) and is computed
directly at cell faces (property {\bf P5}). This latter feature is a
distinctive characteristic that sets $\FLrls$ apart from other length
scale definitions (see Table~\ref{properties_Delta}). Therefore, it
requires a couple of very simple modifications respect to the standard
implementation of an eddy-viscosity model (see
Figure~\ref{flowchart_implementation}). Note that although $\FLrls$
has been derived within the framework of a second-order FVM, a similar
derivation could be applied to other existing spatial discretization
approaches. Therefore, the proposed method is not restricted to FVM.

\mbigskip

Additionally, the alternative dissipation-equivalent definition,
$\FLprls$, given in Eq.~(\ref{DeltaPRLS}), has also been
proposed. Like other existing definitions, $\FLprls$ is evaluated at
the cell centers. Furthermore, this length scale depends on the local
flow topology characterized by the gradient of the resolved velocity,
$\G \equiv \nabla \F{\vel}$ (property {\bf P3}). Regarding this,
analytical studies of simple flow configurations indicate the
suitability of the proposed definitions. \bef{The effectiveness of the
  new subgrid length scales has been numerically demonstrated through
  simulations of decaying isotropic turbulence using two different
  codes. Comparisons with Deardorff’s classical length scale show that
  the proposed definitions are significantly more robust to mesh
  anisotropies. Given these results, along with its simplicity, we
  believe the proposed length scale has strong potential for
  application in SGS models, particularly in complex geometries
  involving highly skewed or unstructured meshes.}

\mbigskip

\aft{Numerical results have demonstrated the effectiveness of the
  proposed subgrid characteristic length scales through simulations of
  decaying isotropic turbulence using two different codes. Comparisons
  with Deardorff’s classical definition, $\FLvol$, confirm that both
  new approaches, $\FLprls$ and $\FLrls$, are significantly more
  robust to mesh anisotropies. While both perform well, $\FLrls$
  consistently exhibits greater robustness in highly anisotropic
  configurations. This improved behavior has also been confirmed in a
  turbulent channel flow at $Re_\tau=395$, further supporting its
  applicability in practical LES settings. Given its favorable
  mathematical properties, ease of implementation, and minimal
  computational overhead, we conclude that $\FLrls$ is a reliable and
  broadly applicable choice for SGS modeling in complex geometries,
  including those involving highly skewed or unstructured meshes.}

\mbigskip

\aft{Finally, it is worth mentioning that the selection of the
  appropriate turbulent length scale is of crucial importance for
  hybrid RANS-LES approaches, particularly in the context of DES. As
  demonstrated in our previous research~\cite{PONTRI20AIAA}, a SGS
  model that provides very good GAM properties was unable to guarantee
  a proper solution in the regions of resolved turbulence on
  anisotropic meshes. This is of particular significance for
  aeroacoustics~\cite{DUBRUATRI22AIAA}, where deficiencies in the
  simulation of resolved turbulence dynamics can have a substantial
  impact. The {\it rational length scale} thus emerges as a promising
  candidate for use in DES simulations as well.}


\section*{Acknowledgements}

F.X.T. and J.R. are supported by SIMEX project (PID2022-142174OB-I00)
of the {\it Ministerio de Ciencia e Innovaci\'{o}n} MCIN/AEI/
10.13039/501100011033 and the European Union Next
GenerationEU. Calculations were carried out on the MareNostrum~5-GPP
supercomputer at BSC. We thankfully acknowledge these institutions.

\end{document}